\documentclass[runningheads]{llncs}
\usepackage[T1]{fontenc}
\usepackage{graphicx}
\usepackage{amssymb}
\usepackage{color}
\usepackage{amsmath}
\usepackage{cleveref}
\usepackage{url}
\usepackage{subcaption}

\newcommand\sql[1]{\texttt{#1}}
\newcommand\eat[1]{}
\newcommand{\repeatthanks}{\textsuperscript{\thefootnote}}

\newcommand{\join}{\ensuremath{\mathbin{\Join}}}

\usepackage{xcolor}
\usepackage{algpseudocode}
\usepackage{listings}
\begin{document}
\title{CrypQ: A Database Benchmark Based on Dynamic, Ever-Evolving Ethereum Data}
\titlerunning{CrypQ}
%
\author{Vincent Capol\inst{1}\thanks{These authors have equal contributions and their names are sorted alphabetically.}\orcidID{0009-0000-3501-4814} \and
Yuxi Liu\inst{1}\repeatthanks\orcidID{0009-0004-4115-1427} \and
Haibo Xiu\inst{1}\repeatthanks\orcidID{0009-0006-0012-2368} \and Jun Yang\inst{1}\orcidID{0000-0003-0604-6790}}
\authorrunning{Capol et al.}
%
\institute{Duke University, Durham NC 27708, USA\\
\email{\{vincent.capol,yuxi.liu,haibo.xiu\}@duke.edu}\email{,junyang@cs.duke.edu}}
\maketitle              
\begin{abstract}
Modern database systems are expected to handle dynamic data whose characteristics may evolve over time.
Many popular database benchmarks are limited in their ability to evaluate this dynamic aspect of the database systems.
Those that use synthetic data generators often fail to capture the complexity and unpredictable nature of real data,
while most real-world datasets are static and difficult to create high-volume, realistic updates for.
This paper introduces CrypQ, a database benchmark leveraging dynamic, public Ethereum blockchain data.
CrypQ offers a high-volume, ever-evolving dataset reflecting the unpredictable nature of a real and active cryptocurrency market.
We detail CrypQ's schema, procedures for creating data snapshots and update sequences, and a suite of relevant SQL queries.
As an example, we demonstrate CrypQ's utility in evaluating cost-based query optimizers on complex, evolving data distributions with real-world skewness and dependencies.

\end{abstract}

\section{Introduction}
\label{sec:introduction}

Modern database systems are expected to handle dynamic data whose characteristics may evolve over time. Being able to offer consistent performance in this dynamic setting presents significant challenges.
For example, one challenge arises in cost-based query optimization.
A query has many equivalent query plans that yield the same result, but with vastly different execution costs.
Without executing these plans, cost-based optimizers predict their costs and select one with the lowest cost.
These predictions are often based on compact summaries or models reflecting the current data distribution.
As tuples are inserted or updated, such representations may become outdated and thus lead to poor cost predictions and suboptimal, potentially catastrophic, choices of query plans.
An effective database system must find ways to overcome this issue, especially when faced with challenging query workloads and high volumes of updates on unpredictable data with complex distributions.

A database benchmark must evaluate the ability of a database system to perform in this regard.
However, as discussed in the related works section,
many popular database benchmarks do not meet this need.
Benchmarks utilizing synthetic data often fail to capture the complexity and unpredictability of real data,
while most real-world datasets are static and difficult to create high-volume, realistic updates for.

This paper introduces \emph{\textbf{CrypQ}} (\url{https://github.com/dukedb-crypq}~\cite{our-benchmark})
a public database benchmark leveraging dynamic, public Ethereum\footnote{\url{https://ethereum.org/en/what-is-ethereum/}} blockchain data.
Ethereum, conceived by Vitalik Buterin in 2013, is a decentralized blockchain that has gained considerable traction, recording 1,420,187 daily active addresses at its peak in late 2022,\footnote{\url{https://etherscan.io/chart/active-address}} and 1,961,144 daily transactions in January 2024.\footnote{\url{https://etherscan.io/chart/tx}}
Enabled by the public nature of Ethereum's ledger,
which records all transactions that occurred on the blockchain,
this large pool of Ethereum data is available for anyone to see. 

Using an external tool, Google BigQuery directly obtains data from a node in the Ethereum network~\cite{ethereum-bigquery},
exporting it as a public dataset on a daily basis.
This dataset allows an ever-increasing amount of Ethereum data to be accessed easily through BigQuery.
Utilizing a modified version of this dataset, we are able to offer a new database benchmark with several distinct advantages:

\begin{enumerate}
    \item Dynamic data: Our benchmark naturally incorporates dynamic changes, including data insertions, updates, and (optionally) expiration, eliminating the need for synthesizing such workloads.
    \item Realistic distributions: Our dataset naturally exhibits a wide array of challenging characteristics, including time-varying correlations and skewness. Therefore, our workloads can stress-test aspects of the database systems such as query optimization in a realistic manner, which is difficult to do with synthetic data.
    \item Scalability: The source of our data is large and grows on a continual basis, allowing our benchmark to scale easily and accommodate samples of different sizes. Beyond scaling simply in size, new data offers additional novelty and unpredictability, which are not possible with existing benchmarks that synthesize data from fixed distributions.
\end{enumerate}

In the rest of this paper, we first cover related work on database benchmarking.
We then describe the CrypQ benchmark in more detail, starting with some background on Ethereum to help understand our schema and workload.
We then present some example uses of CrypQ to illustrate its usefulness as a benchmark for studying database performance,
focusing specifically on cardinality estimation and plan selection in PostgreSQL.

\section{Related Work}
\label{sec:related}

\paragraph{Relational Data Management Systems Benchmarks}

We compare CrypQ with five popular benchmarks in this category, summarized in \Cref{tab:datasets}.

\begin{table}[t]\small
\centering
\begin{tabular}{l||c|c|c|c|c|c}
\hline
& {TPC-DS} & {DSB} & {JOB} & {STATS} & {Stack} & \textbf{CrypQ} \\ \hline\hline
\emph{Uses real-world data?} & \textbf{×} & \textbf{×} & \checkmark & \checkmark & \checkmark & \checkmark \\ \hline
\emph{Can be scaled up easily?} & \checkmark & \checkmark & \textbf{×} & \textbf{×} & \textbf{×} & \checkmark \\ \hline
\emph{Provides data updates?} & $-^\dagger$ & $-^\dagger$ & \textbf{×} & \textbf{×} & \checkmark$^\ddagger$ & \checkmark \\ \hline
\emph{Features time-varying data distribution?} & \textbf{×} & \textbf{×} & \textbf{×} & \textbf{×} & \checkmark & \checkmark \\ \hline
\end{tabular}
\vspace{0.5ex}
\caption{Summary of comparison of relational database benchmarks.
\itshape $\dagger$: periodic refreshes only; $\ddagger$: monthly updates only.}
\label{tab:datasets}
\end{table}

TPC-DS~\cite{tpcds}, modeled as a data warehouse of a typical retail product supplier, uses synthetically generated data for its 24 tables, allowing it to scale easily. The benchmark has 99 parameterized query templates, as well as 12 data maintenance operations that reflect tasks likely to be performed in practice. These tasks range from complex decision support problems to periodic data refreshes. In order to generate data that scales easily, its data generators make simple assumptions that are not time-varying, which systems can take advantage of easily, making this benchmark easier than real-world scenarios, as many recent experimental papers have noted.

DSB~\cite{ding2021dsb} is another synthetic benchmark, offering itself as an enhanced version of TPC-DS. Unlike the static query workload in TPC-DS, DSB is able to generate a dynamic query workload. DSB also contains more challenging query templates, including many-to-many joins, inequality joins, and cyclic joins. In terms of data, DSB introduces more complex distributions and correlations for both single columns and multi-dimensional data within or across tables. However, it still synthetically generates data from three types of exponential distribution, without modeling distribution shifts across refreshes.

JOB~\cite{leis2015good} is a well-known benchmark for evaluating cardinality estimation and join ordering. JOB has 21 tables, utilizing real-world data from the Internet Movie Database (IMDB). It provides 33 query templates and 113 query instances generated from these templates. While IMDB provides real, non-uniform data distributions and JOB offers diverse join relationships, the static nature of IMDB restricts its usefulness when analyzing adaptability and the performance of cardinality estimators in dynamic settings.  Even though one can potentially obtain an update workload by simulating ``time travel'' on historical data in IMDB, the nature of its domain means that the rate of data change would be very slow.

STATS\textcolor{red}~\cite{han2021cardinality} is based on real-world data extracted from the Stats StackExchange. Similar to JOB, this benchmark is used to assess cardinality estimation and end-to-end performance using a static dataset. STATS consists of 8 tables, and the benchmark STATS-CEB includes 146 queries generated from 70 query templates. STATS-CEB's authors suggested their benchmark offers larger scale, more complex distribution, richer join schemas, and more diverse workloads in comparison to JOB. However, like JOB, STATS is limited by its static snapshot of the Stats StackExchange data.

Stack~\cite{marcus2021bao} is a real-world dataset initially introduced in the context of learned query optimizers. It comprises 100GB of questions and answers from StackExchange websites and includes 5,000 selected queries. Although Stack partitions the data by month to simulate data drift, it does not define a fine-grained data update workload. In contrast, our benchmark, CrypQ, offers update workloads that are far more fine-grained and demanding in terms of rate and volume.

\paragraph{Stream Data Management Systems Benchmarks}
More dynamic datasets can be found in the context of benchmarks for stream data management systems, which specialize in real-time processing as data continuously arrive.

Linear Road~\cite{linear_road} simulates a toll system for the motor vehicle expressways of a fictional metropolitan area.
Coordinates of each vehicle are reported every 30 seconds to determine pricing based on the amount of vehicles utilizing the path at a given point in time. Queries determine toll prices, get account balances, view toll history, etc. While the dynamic aspect of this benchmark is compelling, the data is synthetic, and the queries tend to be simple from the perspective of relational optimization.

Yahoo!\ Cloud Serving Benchmark (YCSB)~\cite{ycsb} is a framework/tool to facilitate the comparisons of NoSQL databases. Their workload is based on 4 built-in distributions (uniform, zipfian, latest, and multinomial), with parameters such as update heavy, read heavy, etc. The records are tuples with a fixed number of fields, filled with random fixed-length ASCII strings. Overall, YCSB data and query workloads are very synthetic.

Real-world data from the Internet of Things (IoT) by a real industrial sensor dataset is utilized by~\cite{hendawi2019benchmarking} to benchmark stream data management systems. Their dataset (550 files, 100 GB) is highly unstructured and sparse, and their workloads focus on evaluating parallel execution. Although this dataset is real, the benchmark was not intended for evaluating relational query optimization,
and it would be non-trivial to prepare the data and design appropriate query workloads for that purpose.

SmartBench~\cite{smartbench} is a benchmark that also utilizes IoT data. Besides simple reads and writes, SmartBench proposes more complex queries including joins and aggregations and makes a comprehensive comparison among relational database systems, time-series database systems, document stores, and other NoSQL database systems. However, their queries are designed for evaluating the performance of different database technologies on supporting analytics for IoT data, and hence not representative of traditional relational queries.

ESPBench~\cite{espbench} combines data from two domains: sensor data from manufacturing equipment and business data based on TPC-C benchmark~\cite{tpcc}. One of ESPBench's main contributions is that it fully covers the core functionalities of stream data management system proposed by~\cite{gray1992benchmark}. Combining real sensor data with TPC-C makes the benchmark more comprehensive, but the synthetic nature of TPC-C makes the combined workload less compelling.

Overall, the stream data management benchmarks, although dynamic by definition, require non-trivial effort to retrofit them to the relational setting,
and most of them still rely on synthetic data generation.

\section{Our Benchmark}
\label{sec:benchmark}

\paragraph{Background and Database Schema}

\begin{figure}[h]
\centering
\includegraphics[width=\linewidth]{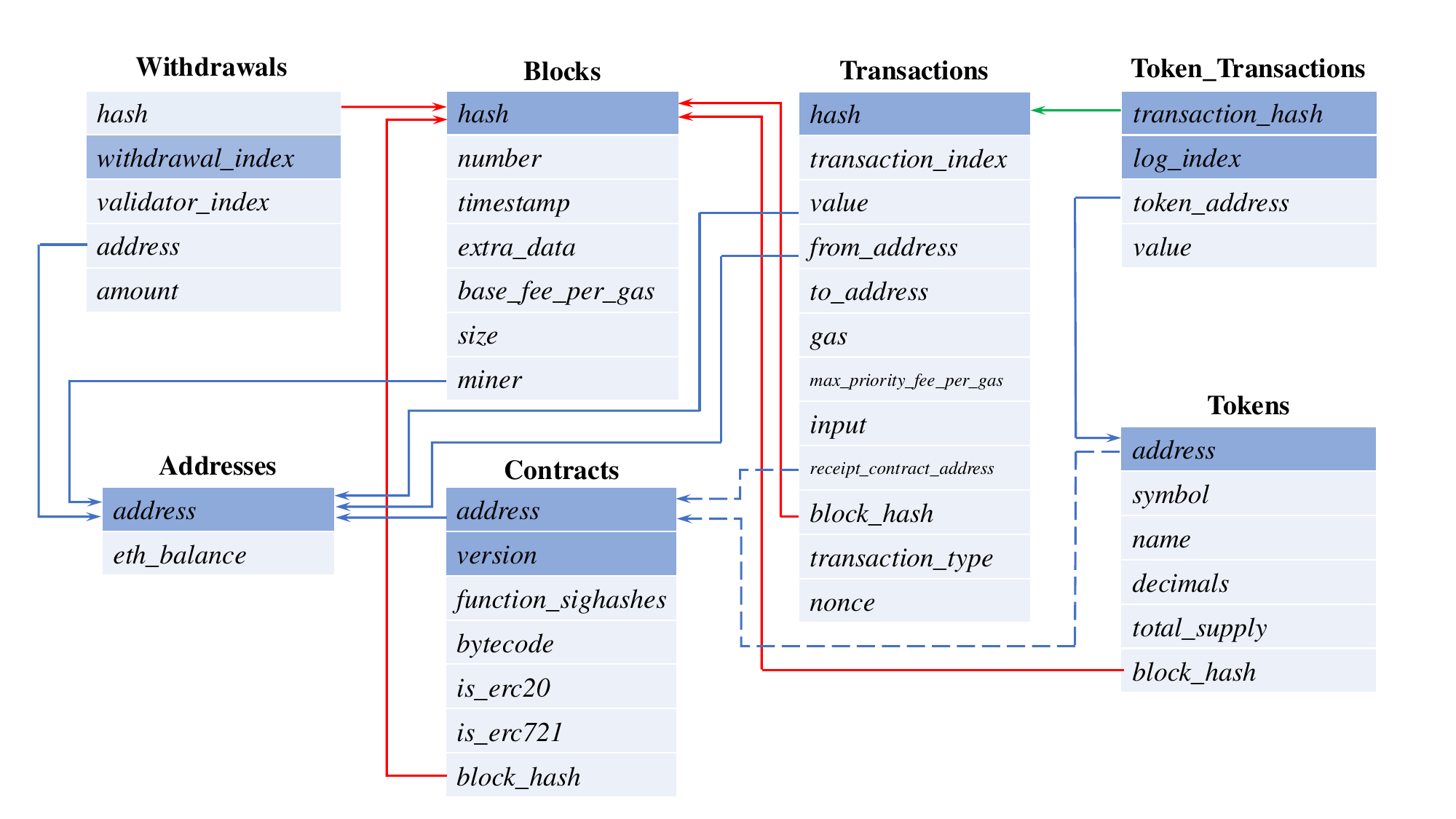}
\caption{CrypQ schema.\itshape
Components of the primary key for each table are shaded;
\sql{UNIQUE} keys are not marked.
Arrows go from foreign keys to the primary keys they reference;
two lines are dashed because they do not reference primary keys.}
\label{fig:schema}
\end{figure}

\Cref{fig:schema} shows the database schema for CrypQ.
As mentioned in \Cref{sec:introduction}, Google maintains all data from Ethereum in BigQuery~\cite{ethereum-bigquery},
so our database schema is quite similar.
We will highlight some differences when discussing data preparation later.
Here, we provide a high-level overview of the schema along with some background on blockchain and Ethereum;
more detailed descriptions of the schema can be found in the (extensively commented) file \texttt{create.sql} in the benchmark.

Due to the nature of blockchains, the \emph{Blocks} table is at the center of the schema; most other tables reference it.
Each row in \emph{Blocks}, identified by its \emph{hash}, represents a block in the Ethereum blockchain.
Each block extends the chain by applying a one-way hash function to the previous block's data.
New blocks are added to the blockchain at the typical rate of several per minute.

The current consensus model used by Ethereum is \emph{proof-of-stake}, which was changed from \emph{proof-of-work} in September 2022.
In the traditional proof-of-work consensus model, which is still used by Bitcoin,
a number of \emph{validators} compete to perform the expensive trial-and-error computational task of finding the new block's \emph{nonce},
which was used in generating its hash.
The first validator to correctly guess this number is rewarded --- in the case of Ethereum, with some amount of \emph{Ether}, the currency used by Ethereum;
then, the new block is officially connected, with its \emph{miner} being this validator, identified by an \emph{address} in the \emph{Addresses} table.
As this method is energy-intensive, Ethereum made the transition to proof-of-stake,
where a major difference is that instead of having all validators compete to validate a new block, one is randomly selected for the task.
In order to have the chance to be selected, validators must lock up (also known as ``stake'') 32 Ether.
Successfully completing the validation is still rewarded, while malicious behavior is punished by confiscating the amount staked.\footnote{\url{https://ethereum.org/en/developers/docs/consensus-mechanisms/pos/}}
Whenever a validator withdraws some amount of their staked Ether, this information is captured by the \emph{Withdrawals} table.
There is no data prior to September 2022 in \emph{Withdrawals}, as staking did not exist.

The \emph{Addresses} table tracks all accounts (including the aforementioned validators) in the Ethereum ecosystem, along with their current Ether balances.
The \emph{Transactions} table tracks transactions, which are recorded within blocks in the blockchain.
Once added to a block, they remain immutable.
There are typically around 100 transactions per block, but this number can vary drastically.
Each transaction is between pair of addresses --- a sender (\emph{from\_address}) and a receiver (\emph{to\_address}) ---
and involves the transfer of some amount of Ether (\emph{Transactions.value}) and/or some \emph{tokens},
which are forms of assets supported by Ethereum and tracked by the \emph{Tokens} table.
If the transaction involves the latter,
entries of the \emph{Token\_Transactions} table link the transaction with the tokens it involves,
and record the quantity of each token transacted.

A big draw of Ethereum is that it is not just a decentralized payment network, but it also allows users to build and deploy decentralized applications.
Such applications employ \emph{smart contracts}, which are computer programs living on the Ethereum blockchain.
In fact, tokens are minted and managed by these contracts.
These contracts are identified by their addresses,
but the same contract can have multiple versions over time, with possibly different set of supported functions,
all tracked by the \emph{Contracts} table.
In general, rows in \emph{Addresses} include both user-controlled accounts as well as smart contract accounts,
and a transaction can involve both types of addresses.

\paragraph{Data Preparation and Update Workload}

The availability of all Ethereum data in Google BigQuery~\cite{ethereum-bigquery} makes it easy to prepare data for CrypQ,
without having to directly interact with the Ethereum blockchain.
However, the dataset (\texttt{bigquery-public-data.crypto\_ethereum}) can be too large to work with for most purposes:
e.g., the \emph{Transactions} table alone contains about 2.4 billion rows and is close to 2TB uncompressed at the time of this writing;
the data is still growing by day.
Therefore, we provide instructions and scripts~\cite{our-benchmark} to extract a data \emph{slice} of the desired size to populate our schema.
A slice contains a number of consecutive blocks in the blockchain ordered by time.
We extract all \emph{Blocks} rows within the corresponding time window,
along with data in other tables that are related to these blocks.%
\footnote{Note that it is possible that some token or contract was created in a block before the current slice.
In that case, we set its \emph{block\_hash} to \sql{NULL} instead of including the referenced block in the current slice.}
Depending on the size of the dataset that a user wishes to obtain, the number of blocks to extract can be adjusted.

We perform some post-processing to transform and clean up the extracted data.
For example, when transform data to conform to our relational schema,
we convert binary data such as hashes and addresses, which are represented as character strings in the BigQuery dataset,
into much more efficient byte representations.
We also extrapolate the \emph{eth\_balance} values of addresses at the end of the extracted slice to be consistent with extracted data,
since the BigQuery dataset does not track balance history.
For additional details, see the \texttt{load.sql} script in our benchmark.
To get a sense for how to control the benchmark database size using the number of extracted blocks,
we note that for the CrypQ slice with 2000 blocks numbered between 19,005,000 and 19,006,999,
the BigQuery export files take close to 1.2GB,
and when finally loaded into PostgreSQL, the database size is about 510MB.

Besides bulk-loading a slice of the dataset,
CrypQ also supports extracting an update workload from the Ethereum blockchain, in the form of a sequence of database modification statements.
On a blockchain, the addition of a new block is what triggers the recording of many other events.
Given a slice of data, say, pertaining to blocks numbered between 19,005,000 and 19,006,999,
we provide a script \texttt{gen\_updates.sh} that generates a load file for creating the initial database state,
e.g., with the first 1000 blocks between 19,005,000 and 19,005,999,
as well as a list of (sequentially named) \texttt{upserts-*.sql} files containing modification statements, batched according to a user-specified granularity, for advancing the database state over time.
Continuing with the example, if the granularity is set to one block at a time,
then 1000 batches will generated from the 2000 blocks extracted, one for each of the blocks between 19,006,000 and 19,006,999.
For each batch, the modification statements will insert new blocks plus withdrawals, transactions, and associated token transactions on these blocks,
as well as any referenced tokens, contracts, or addresses that are previously unseen.
They will also update balances in \emph{Addresses} as a result of the new transactions.
By default, the vast majority of the workload consists of \sql{INSERT} statements, which increases the database size over time.
Optionally, the user can keep the database size roughly constant by asking CrypQ to create an update workload
that essentially maintains a moving slice over the blockchain, with a constant number of blocks in the slice.
Under this option, before adding data in a new batch,
a batch of \sql{DELETE} statements (in \sql{expire.sql}) will first expire data pertaining to the oldest blocks in the database.

A typical use case for the CrypQ update workload would be as follows.
For the running example, the user first bulk-loads the initial database state, e.g., with 1000 blocks between 19,005,000 and 19,005,999.
The user then executes the list of \texttt{upserts-*.sql} files (and optionally, \sql{expire.sql} before each) in order.
Assuming each batch is one block,
after executing \sql{expire.sql} and the first \texttt{upserts-*.sql} file,
the database will contain the 1000 blocks between 19,005,001 and 19,006,000.
Between the update batches, the user can run the query workload or perform other experiments.
If the user wishes to simulate the update workload in a real-time fashion,
a custom workload injector can look for the \emph{timestamp} for the next block to be inserted in an \texttt{upserts-*.sql} file,
and apply that file at the appropriate time.

\paragraph{Query Workload}

Our query workload currently includes 10 queries with varying complexity.
They include single-block queries with up to 5 joins (with and without grouping and aggregation),
multi-block queries with subqueries (with and without correlation and/or aggregation),
as well as queries with advanced features such as lateral joins, window functions, and recursive common table expressions.
The 10 queries are available at~\cite{our-benchmark} and summarized in \Cref{table:queries_charateristics}.
The goal is to present challenges to the query optimizer
while addressing real-world questions relevant to users interested in the Ethereum ecosystem.
We welcome community contributions to enhance this query workload.
Below we show three queries as examples.

\newcommand{\features}[4]{\multicolumn{1}{|c|}{\vbox{\scriptsize\hbox{\strut #1}\hbox{\strut #2}\hbox{\strut #3}\hbox{\strut #4}}}}
\begin{table*}[t]
    \begin{center}
        \begin{tabular}{r||r|r|r|r|r|r|r|r|r|r|}
               & $Q_1$ & $Q_2$ & $Q_3$ & $Q_4$ & $Q_5$ & $Q_6$ & $Q_7$ & $Q_8$ & $Q_9$ & $Q_{10}$   
             \\\hline\hline
             Features$^\dagger$ & \features{str}{}{}{} & \features{cte}{}{}{} & \features{cte}{aggr}{}{} & \features{aggr}{sub}{}{} & \features{cte}{sub}{aggr}{win}
                & \features{aggr}{c-sub}{}{} & \features{cte}{c-sub}{aggr}{set} & \features{aggr}{}{}{} & \features{cte}{lateral}{c-sub}{str} & \features{r-cte}{aggr}{sub}{set} 
             \\\hline
             \# table instances$^\ddagger$ & $5$ & $3$ & $7$ & $3$ & $7$
                & $3$ & $13$ & $3$ & $5$ & $10$
             \\\hline
             Output cardinality & & & & & & & & & &\\
             $S_1$ & 1 & 17 & 1 & 1 & 35 & 0 & 48 & 68 & 0 & 8\\
             $S_2$ & 1 & 23 & 1 & 1 & 50 & 4,039 & 64 & 890 & 357 & 13\\
             $S_3$ & 1 & 13 & 1 & 1 & 50 & 3,666 & 64 & 2,100 & 246 & 3\\
             $S_4$ & 1 & 6 & 1 & 1 & 50 & 3,966 & 32 & 143 & 333 & 61\\
             \hline
             Response latency$^\S$ & & & & & & & & & &\\
             $S_1$ & 31 & 38 & 32,113 & 91 & 108 & 2,364 & 1,554 & 232 & 47 & 192\\
             $S_2$ & 31 & 88 & 85,081 & 252 & 302 & 18,102 & 6,376 & 49,171 & 4,956 & 452\\
             $S_3$ & 34 & 62 & 107,760 & 227 & 209 & 16,486 & 5,755 & 20,045 & 5,926 & 833\\
             $S_4$ & 31 & 39 & 127,770 & 315 & 217 & 42,906 & 5,317 & 185,002 & 5,087 & 369\\
             \hline
        \end{tabular}
    \end{center}
    \caption{Summary of the query workload.
    \itshape $\dagger$
        \emph{aggr}: aggregation (with possible grouping);
        \emph{(r-)cte}: (recursive) common table expressions;
        \emph{lateral}: lateral joins;
        \emph{set}: set/bag operations;
        \emph{str}: string functions;
        \emph{(c-)sub}: (correlated) subqueries;
        \emph{win}: window functions.
        $\ddagger$ Number of table instances appearing in all \sql{FROM} clauses, across the entire query, including subqueries and common table expression definitions.
        $\S$ Response latency is measured as the median of 11 repetitions and in ms, using the experimental setup described in \Cref{sec:use}.
        $S_1$ through $S_4$ are 4 different database slices defined in \Cref{sec:use}.}
    \label{table:queries_charateristics}
\end{table*}

The following query captures large transactions conducted by users with significant activities and balances,
which may be of interest to those analyzing the behavior of major stakeholders in the Ethereum ecosystem:
\begin{lstlisting}[language=SQL, basicstyle=\footnotesize\ttfamily]
-- Q1:
SELECT COUNT(*)
FROM Transactions tx, Tokens tk, Token_Transactions tk_tx, Contracts c, Addresses a
WHERE tx.hash = tk_tx.transaction_hash
AND tk_tx.token_address = tk.address
AND tx.to_address = c.address
AND tx.from_address = a.address
AND tx.nonce BETWEEN 2100000 AND 4200000
AND tk_tx.value BETWEEN 1000000000 AND 10000000000
AND tk.name NOT LIKE '%US%'
AND c.is_erc20 = TRUE
AND a.eth_balance >= 25000000000000000;
\end{lstlisting}
\noindent Specifically, $Q_1$ returns the number of token transactions
(excluding those with ``US'' in the token name)
where between $10^9$ and $10^{10}$ of a token were sent to an ERC20 contract
by a user account who has transacted between $2.1$ and $4.2$ millions times in the past
(\emph{t.nonce} records the number of transactions originated so far from \emph{t.from\_address})
and currently has a balance of at least $25\times10^{15}$.

The next query tracks token transactions that amount to a significant portion (at least 0.01\%) of a token's overall supply.
This information can be useful for monitoring suspicious activities, shifts in sentiment, and asset value fluctuations:
\begin{lstlisting}[language=SQL, basicstyle=\footnotesize\ttfamily]
-- Q2:
WITH Temp AS (
    SELECT tk_tx.*, tk.symbol, tk.name, tk.total_supply,
        tk_tx.value * 100 / POWER(10, COALESCE(tk.decimals, 18)) / tk.total_supply AS percentage
    FROM Token_Transactions tk_tx, Tokens tk
    WHERE tk_tx.token_address = tk.address
    AND tk.total_supply <> 0)
SELECT * FROM Temp
WHERE percentage BETWEEN 0.01 AND 100
ORDER BY percentage DESC;
\end{lstlisting}

Our third example tries to understand the usage patterns of Ethereum contracts
by computing the average duration, measured in the number of blocks passed, between the creation of a contract and its first use:
\begin{lstlisting}[language=SQL, basicstyle=\footnotesize\ttfamily, literate={number}{number}{6},]
-- Q3:
WITH Creation(address, version, number) AS (
	SELECT c.address, c.version, b.number
	FROM Contracts c, Blocks b
	WHERE c.block_hash = b.hash),
First_Used(address, version, number) AS (
	SELECT c.address, c.version, MIN(b.number)
	FROM Contracts c, Blocks b, Transactions t
	WHERE (t.from_address = c.address OR t.to_address = c.address)
	AND t.block_hash = b.hash
	GROUP BY c.address, c.version)
SELECT AVG(First_used.number - Creation.number)
FROM Creation, First_used
WHERE Creation.address = First_used.address
AND Creation.version = First_used.version;
\end{lstlisting}

\section{Example Use of Benchmark}
\label{sec:use}

In this section, we show some example uses of CrypQ to illustrate its usefulness as a benchmark for studying database performance,
especially in dynamic settings.
We will focus on cardinality estimation and plan selection in PostgreSQL here.
Cardinality estimation is believed to be one of the major causes of poor query performance,
and supporting accurate and stable cardinality estimation over a dynamically updated database remains an active research area.
Query plan selection is the next step in query optimization,
and the quality of plan choices depends not only on the accuracy of cardinality estimation but also on other factors,
such as cost functions and the plan enumeration algorithm.
Please note this section is not intended as a full-fledged effort for benchmarking PostgreSQL;
instead, our goal is to demonstrate how CrypQ's wide range of data characteristics and its dynamic nature
pose realistic, interesting challenges to query optimization worthy of investigation.
We consider two scenarios.

\emph{Scenario 1} is designed to mirror real-world situations where the underlying data is constantly evolving.
While individual row modifications each have little impact on the overall data distribution,
over time they may shift distributions and impact cardinality estimation and plan selection.
The cost of refreshing summary statistics or models for cardinality estimation must be balanced against potential improvements in plan quality.
To set up this scenario, we extract data about 2000 blocks numbered between 19,005,000 and 19,006,999 (which started in Jan.\ 2024),
and we then apply the update workload preparation procedure described in \Cref{sec:benchmark} to create
an initial database state $W_1$ consisting of the first 1000 blocks, as well as
10 batches of updates each adding data from 100 new blocks while expiring the oldest 100 blocks,
resulting in 10 subsequent database states $W_2$ to $W_{11}$, each with 1000 blocks,
with $W_{11}$ containing the last 1000 blocks extracted.
Overall, each $W_i$ contains 50\% of the extract data, and shares 90\% of its contents with preceding and succeeding states (if any).

\textit{Scenario 2} picks several slices of the Ethereum blockchain data over the course of several years,
where the slices, not necessarily connected as in Scenario~1, exhibit more variations in data distribution.
This scenario is designed to test the adequacy of cardinality estimation and plan selection for a range of distributions.
Even those summary statistics or models are refreshed for each slice,
some data distributions --- still naturally arising in real life --- may be more challenging than others for query optimization.
To set up this scenario, we extracted $4$ slices labeled $S_1$ to $S_4$, each containing 1000 consecutive blocks but selected using different criteria:
\begin{itemize}
    \item $S_1$ (Jan.\ 2020), the period with the lowest daily transaction count in last 5 years;
    \item $S_2$ (Nov.\ 2021), the period with the highest-ever price of Ethereum;
    \item $S_3$ (Jun.\ 2022), the period with the lowest Ethereum price during the last major downturn;
    \item $S_4$ (Jan.\ 2024, same as $W_1$ in Scenario~1), the period with highest-ever daily transaction count.
\end{itemize}

All our experiments are performed on a Linux server with 8 Intel(R) Core(TM) i7-7700 CPU @ 3.60GHz processors and 1TB of disk storage.
PostgreSQL 16.2 is our database system, and we utilize the methodology outlined in~\cite{xiu2024parqo} for injecting cardinality and query plans into the optimizer.

\paragraph{Example Use 1: Cardinality Estimation}

First, we demonstrate how our benchmark can be used to evaluate the performance of cardinality estimation.
For illustration, we simply evaluate PostgreSQL's default single-column histograms;
in general, we could similarly evaluate any other cardinality or selectivity estimator, be it based on histograms or machine learning.

We extract the subqueries involving $1$ to $3$ tables from $Q_1$ in our query workload (\Cref{sec:benchmark}).
We use the table alias names used in a subquery to denote the subquery:
e.g., $\mathit{tx} \join c$ denotes the subquery joining \emph{Transactions} and \emph{Contracts}.
We execute the same subquery over different database states
and record both PostgreSQL's estimated output cardinality and the actual cardinality revealed by query execution.
Additionally, we report \emph{Q-errors}, a measure widely used for capturing the relative accuracy of the estimates~\cite{qerror}.

\begin{figure}[t]
    \begin{subfigure}[]{0.48\linewidth}
        \centering
        \includegraphics[scale=0.25]{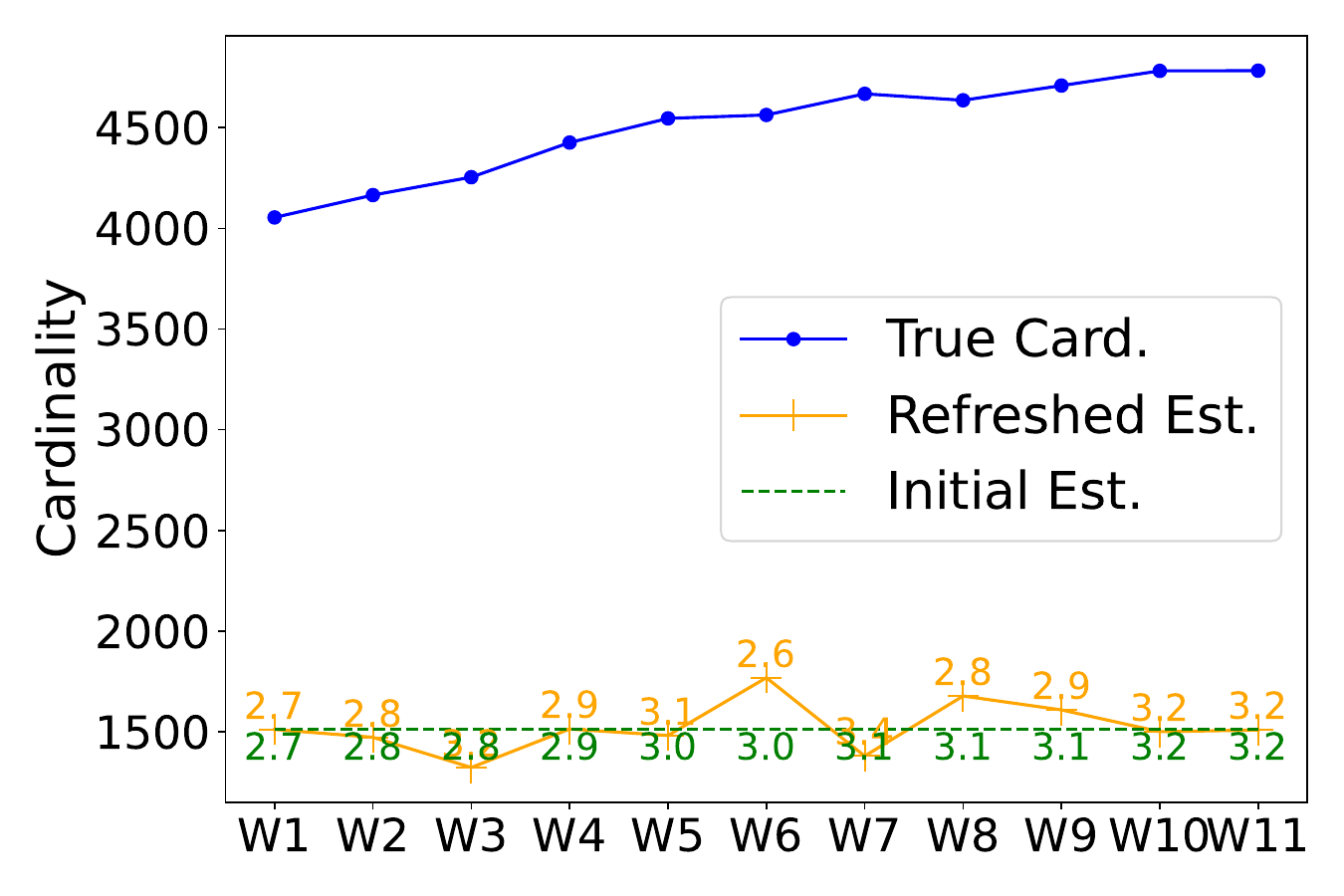}
        \caption{$\mathit{tx} \join a$}
        \label{fig:w:t-join-a}
    \end{subfigure}
    \begin{subfigure}[]{0.48\linewidth}
        \centering
        \includegraphics[scale=0.25]{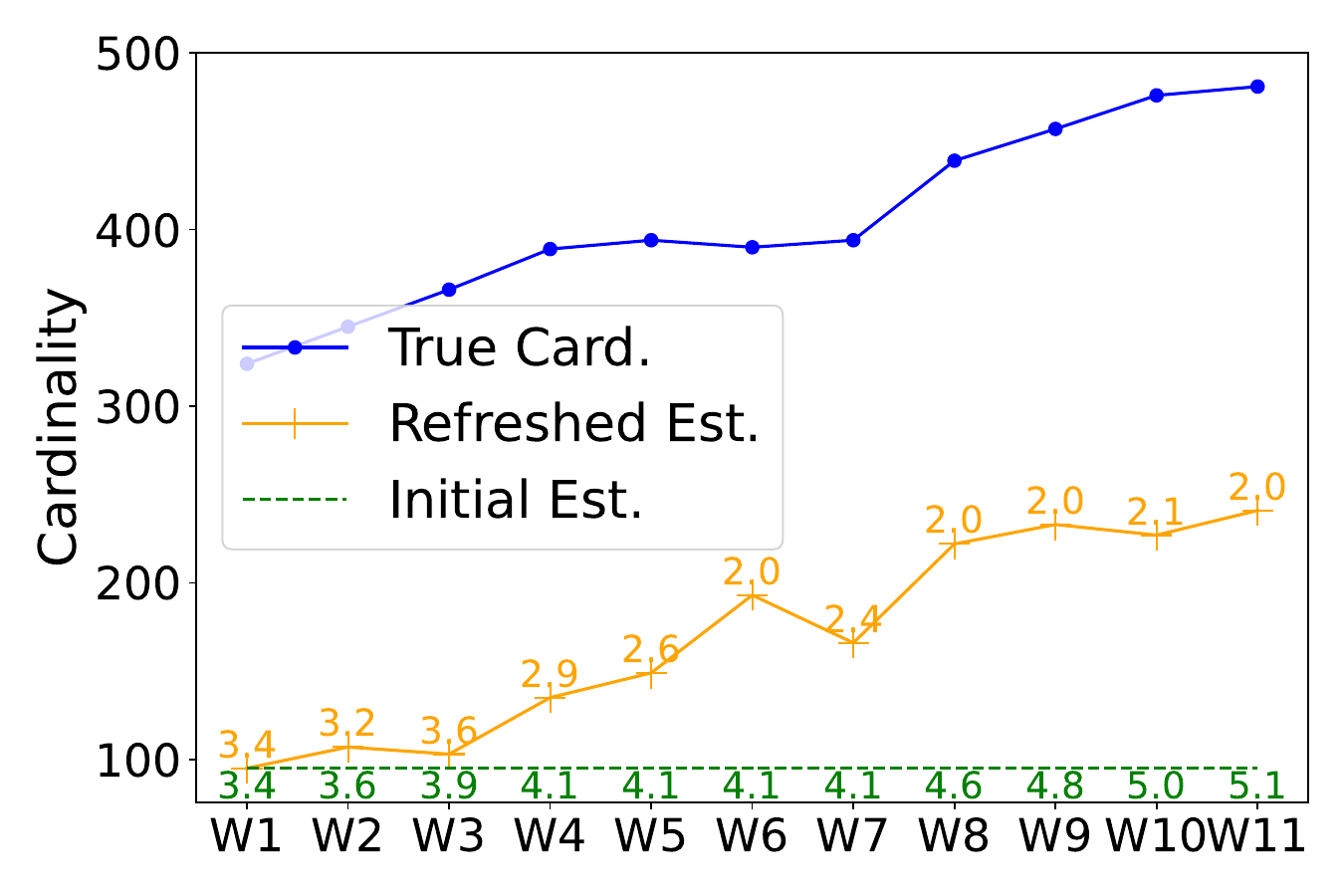}
        \caption{$\mathit{tx} \join \mathit{tk\_tx}$}
        \label{fig:w:t-join-t_t}
    \end{subfigure}
    \\
    \begin{subfigure}[]{0.48\linewidth}
        \centering
        \includegraphics[scale=0.25]{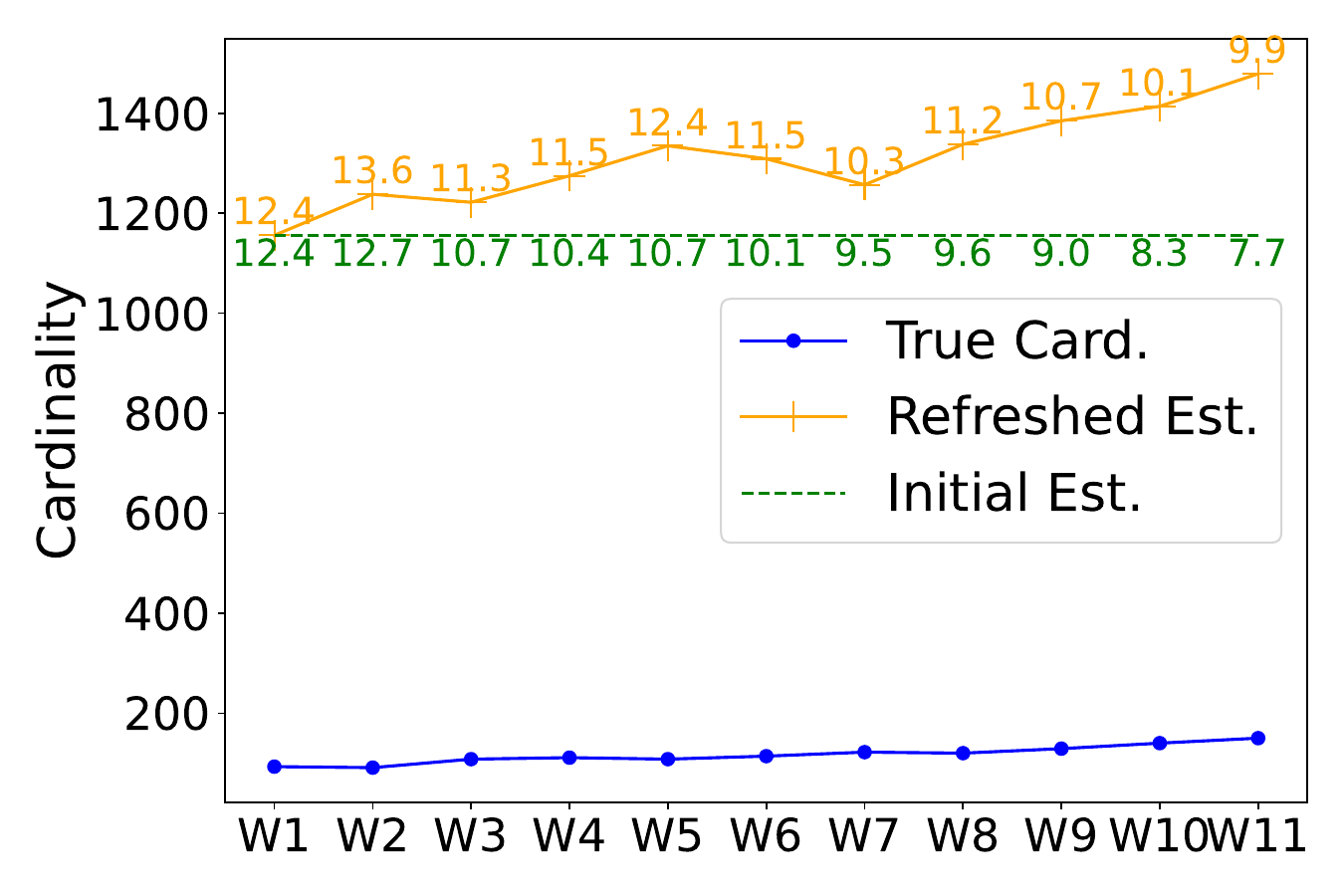}
        \caption{$\mathit{tk} \join \mathit{tk\_tx}$}
        \label{fig:w:token-join-t_t}
    \end{subfigure}
    \begin{subfigure}[]{0.48\linewidth}
        \centering
        \includegraphics[scale=0.25]{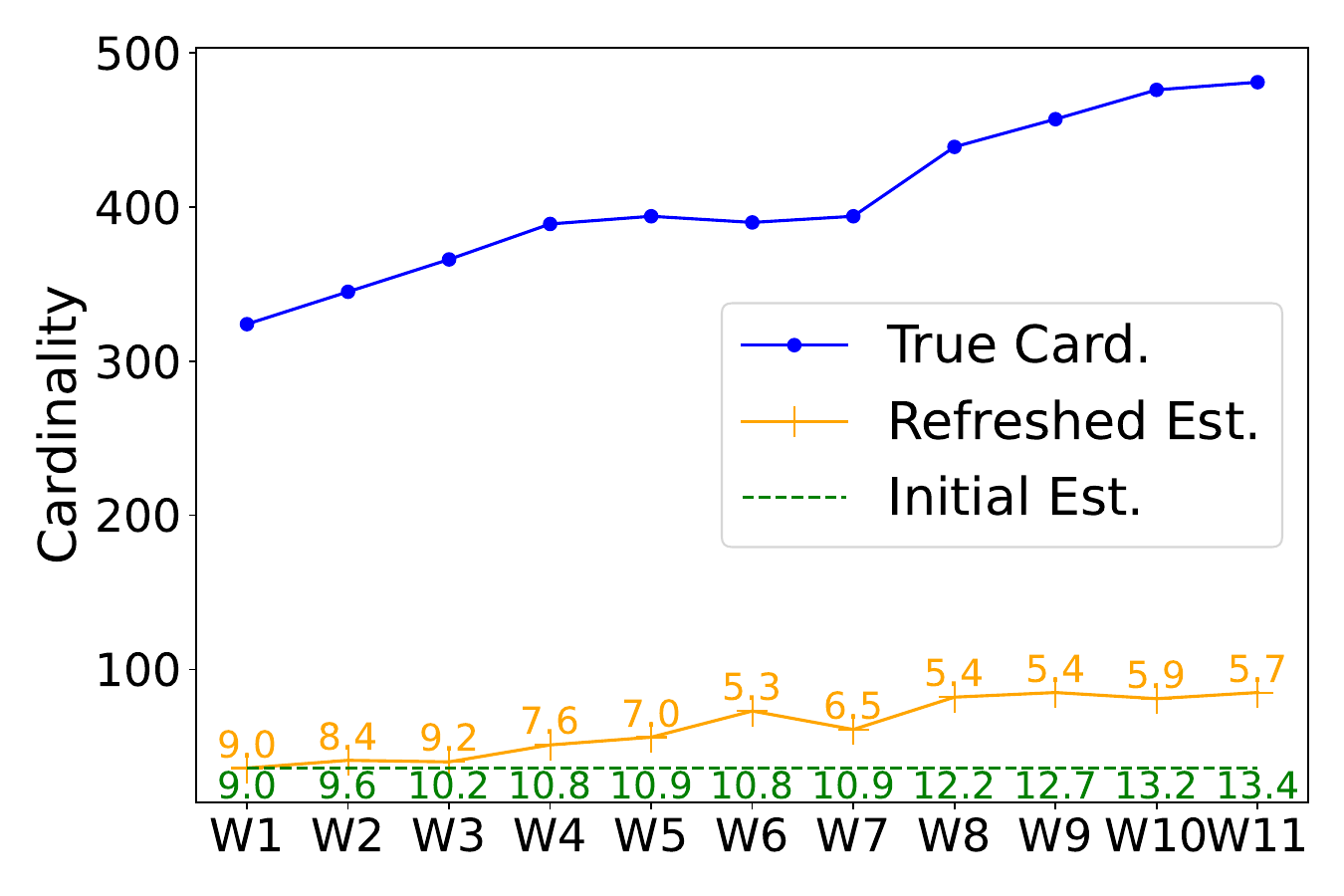}
        \caption{$\mathit{tk\_tx} \join \mathit{tx} \join a$}
        \label{fig:w:t-join-t_t-join-a}
    \end{subfigure}
    \caption{Accuracy of cardinality estimation for Scenario~1.
    \itshape Q-errors with respect to true cardinalities are shown as labels on the lines.}
    \label{fig:w-qerror}
\end{figure}

The results for Scenario~1 are shown in~\Cref{fig:w-qerror}.
For each database state $W_i$, two estimated cardinalities are shown in comparison with the true cardinalities.
The ``refreshed'' estimates are obtained by instructing PostgreSQL to refresh its histograms after every update batch;
the ``initial'' estimates remain the same as those obtained on the initial database state $W_1$,
and reflect the behavior when the cardinality estimator continues to use outdated summary statistics or models despite data updates.
We omit results for subqueries where PostgreSQL's estimates are accurate enough (with Q-error $< 1.01$).
For the four subqueries shown in \Cref{fig:w-qerror}, we see that PostgreSQL estimates can be quite inaccurate (sometimes by an order of magnitude):
underestimating for 3 out of the 4 subqueries and overestimating for one.
When the database evolves from $W_1$ to $W_{11}$, true cardinalities for all subqueries tend to grow.
As long as we instruct PostgreSQL to refresh its histograms following every update batch,
refreshed estimates reflect the same growth trend, so the Q-errors remain relatively stable.
However, without refreshing histograms,
estimates fail to track the trend, so their Q-errors show far more fluctuation:
they become worse over time for 3 out of the 4 subqueries;
they improve for one subquery, but only because the initial estimate was an overestimate.
These results exposes some inadequacy in PostgreSQL histogram-based estimates
and highlights the benefit of refreshing histograms as data changes
in a realistic setting.

\begin{figure}[!h]
    \begin{subfigure}[]{0.32\linewidth}
        \includegraphics[scale=0.18]{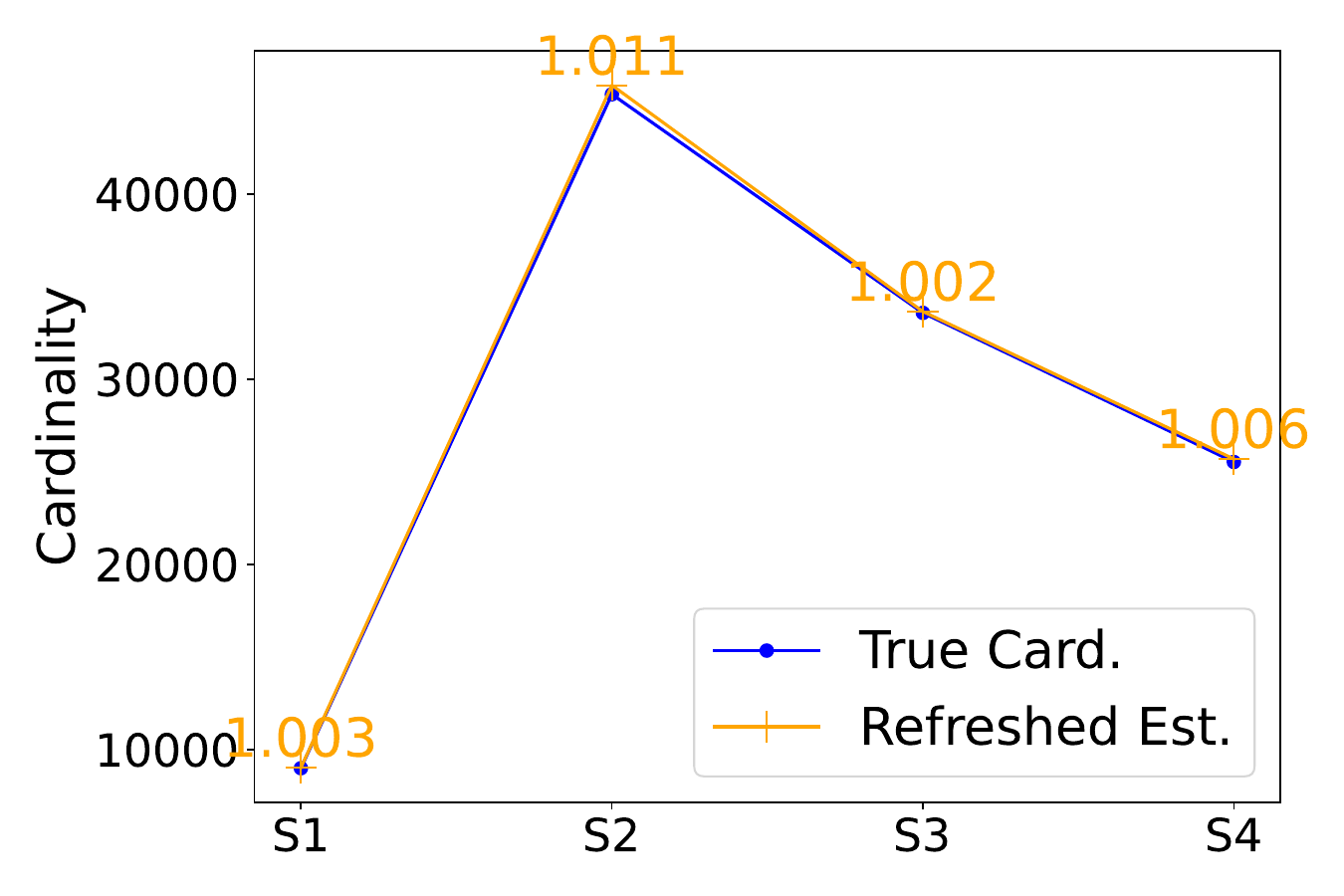}
        \caption{$a$}
    \end{subfigure}
    \begin{subfigure}[]{0.32\linewidth}
        \includegraphics[scale=0.18]{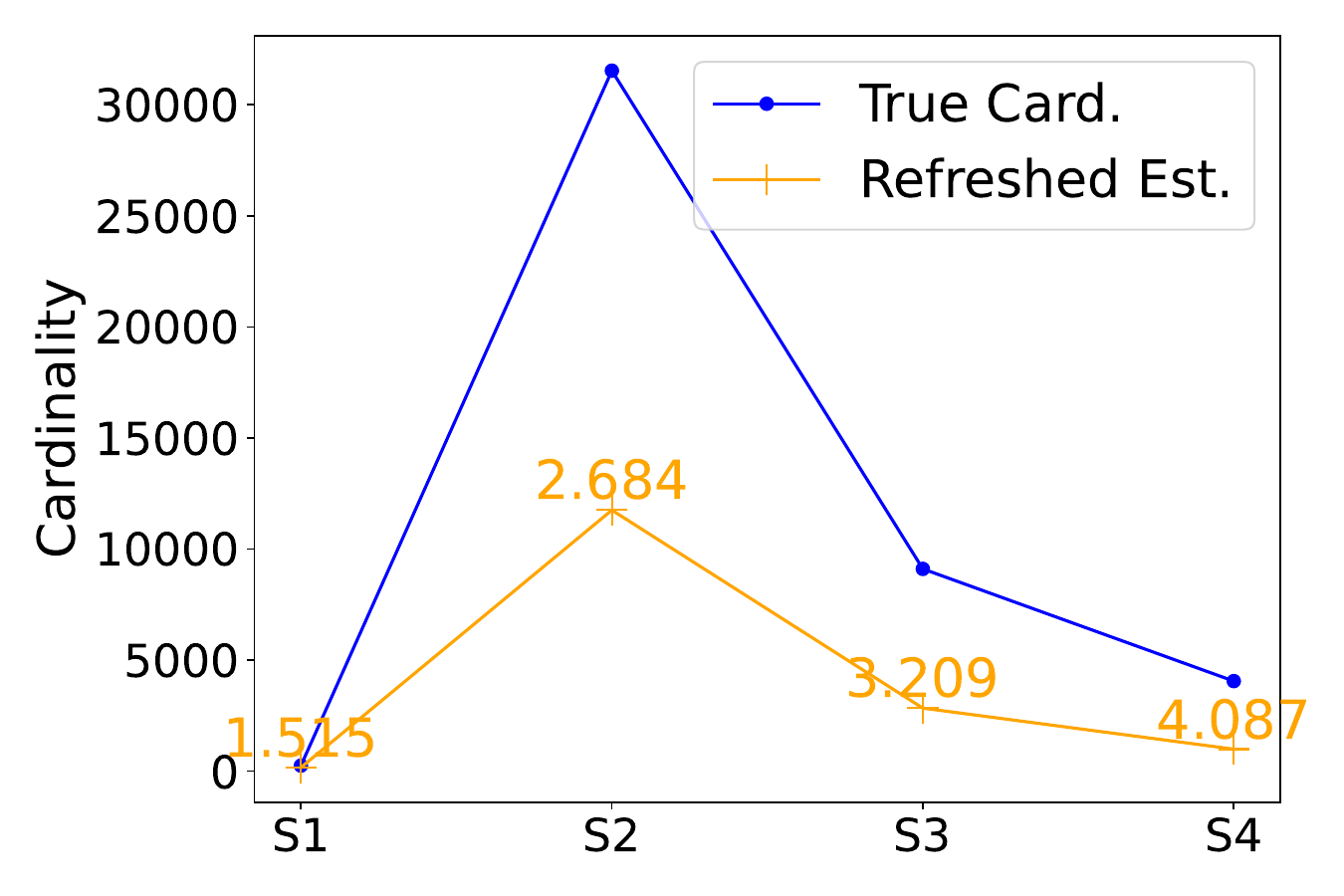}
        \caption{$\mathit{tx} \join a$}
    \end{subfigure}
    \begin{subfigure}[]{0.32\linewidth}
        \includegraphics[scale=0.18]{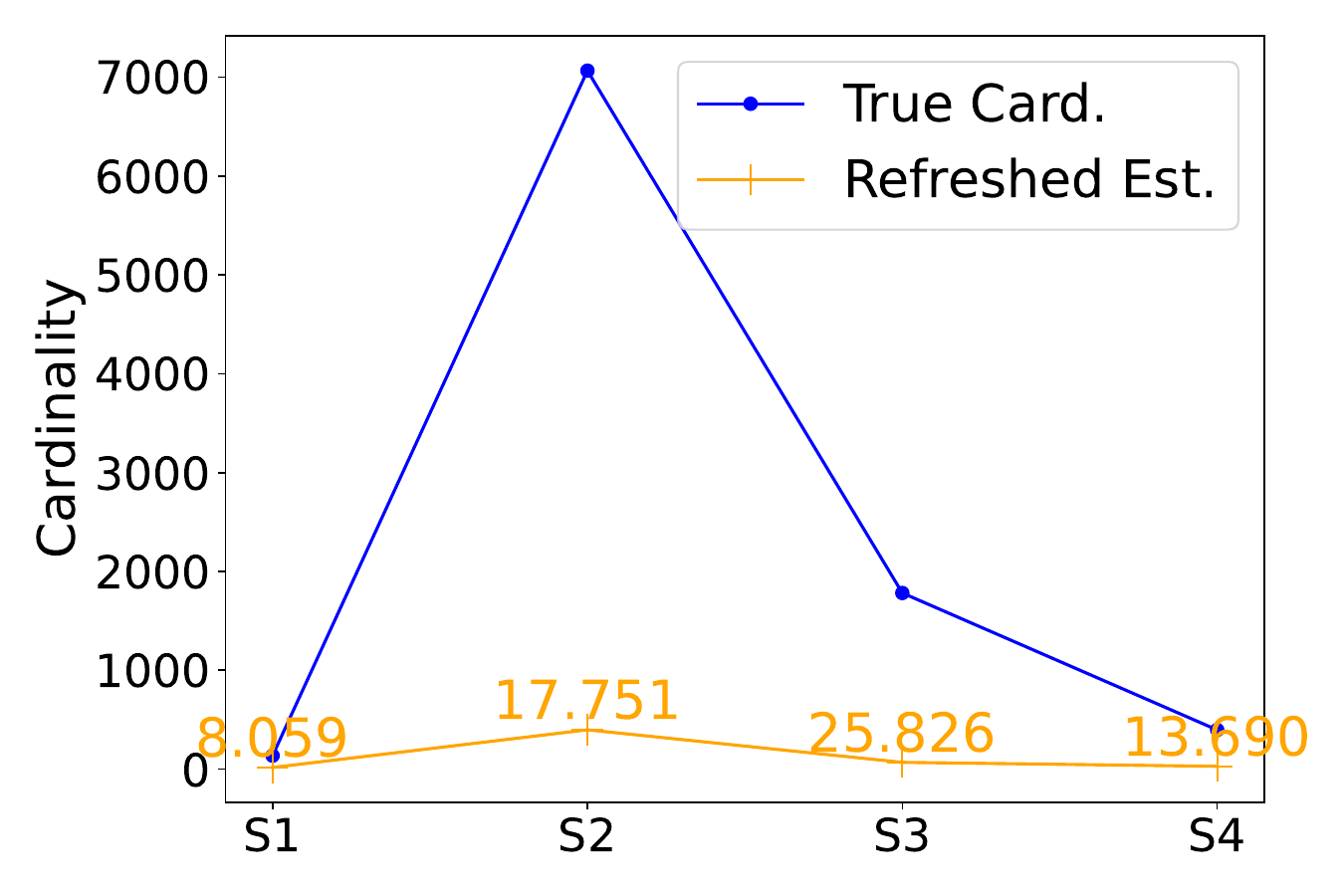}
        \caption{$c \join \mathit{tx} \join a$}
    \end{subfigure}
    \\
    \begin{subfigure}[]{0.32\linewidth}
        \includegraphics[scale=0.18]{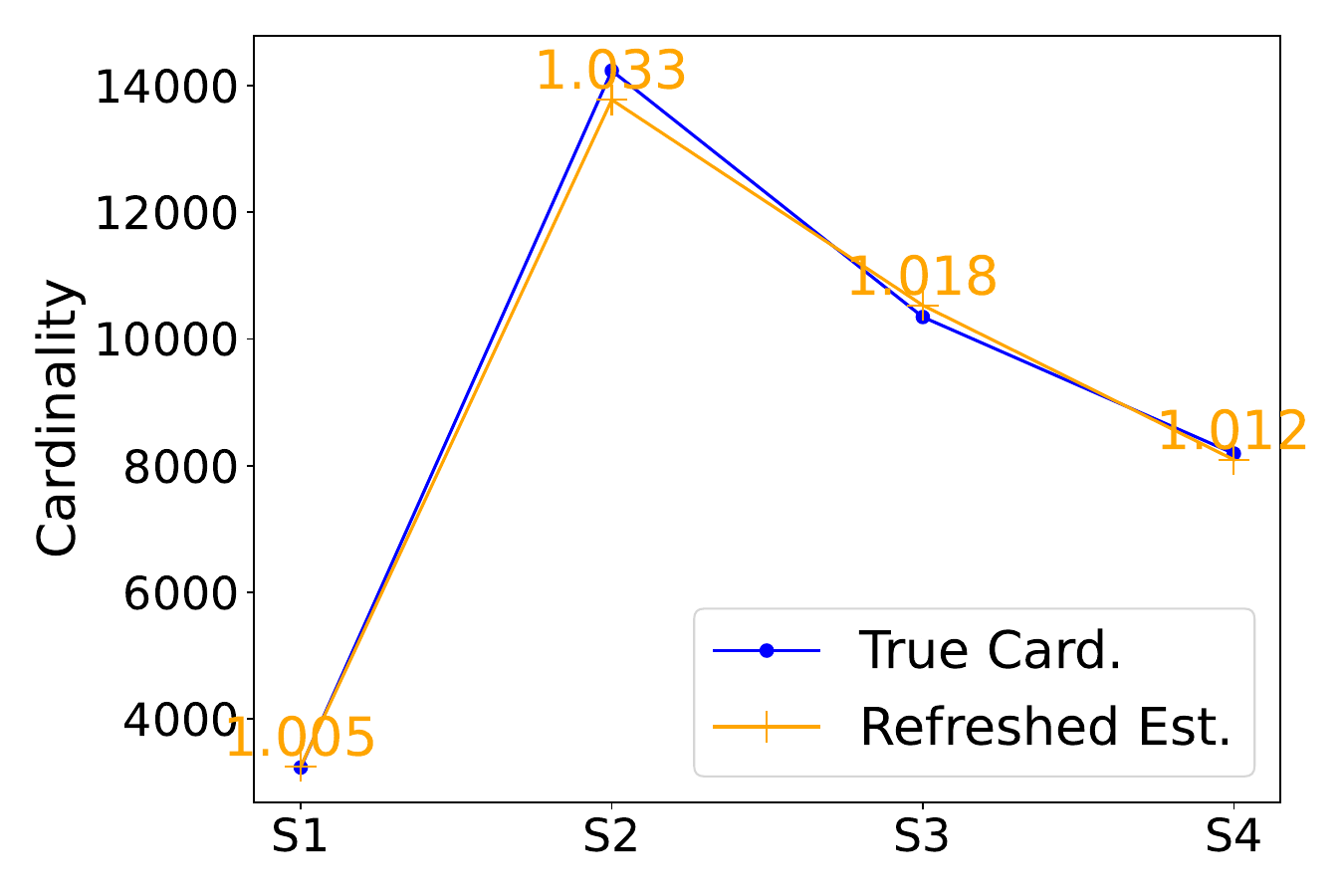}
        \caption{$\mathit{tk\_tx}$}
    \end{subfigure}
    \begin{subfigure}[]{0.32\linewidth}
        \includegraphics[scale=0.18]{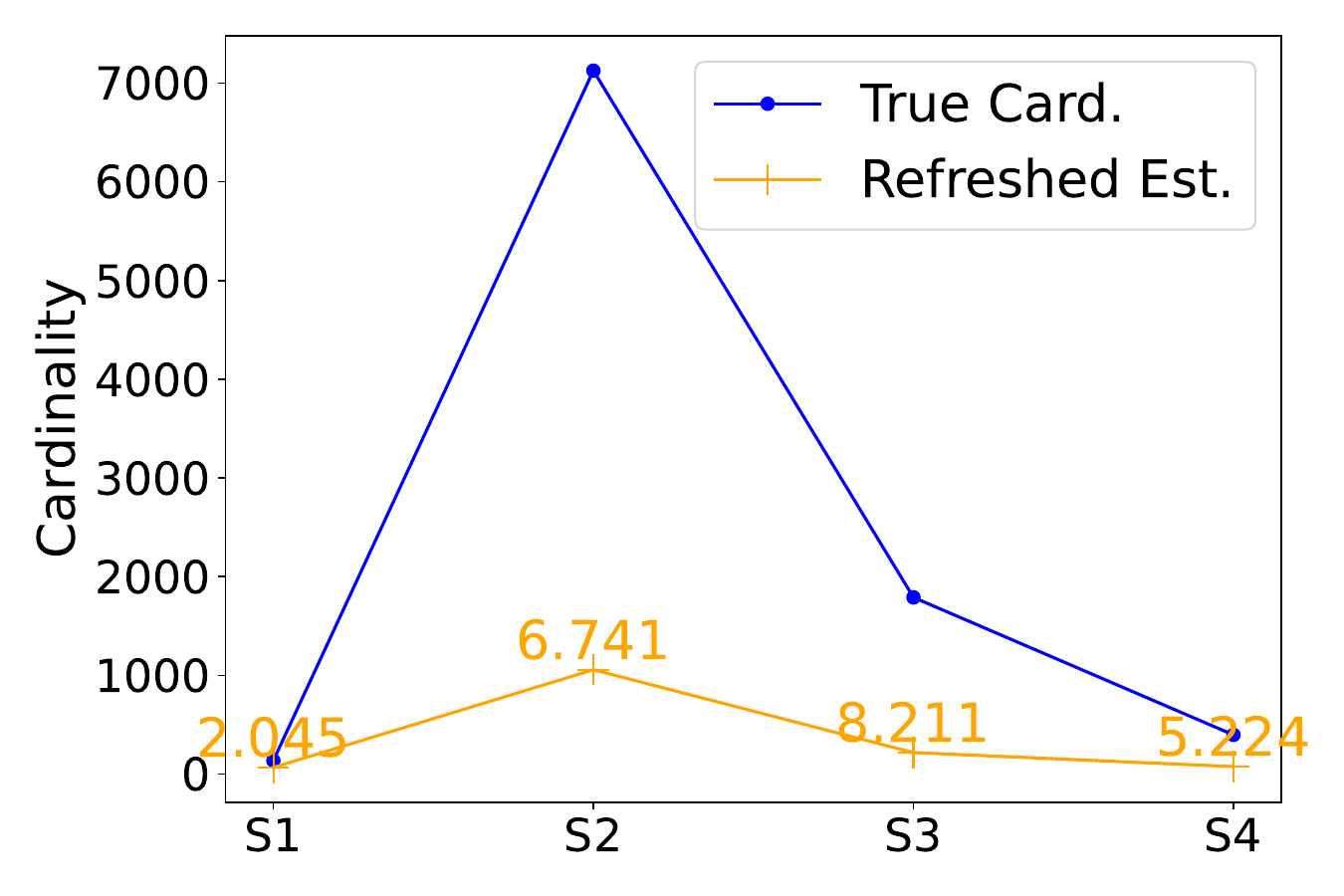}
        \caption{$\mathit{tx} \join c$}
    \end{subfigure}
    \begin{subfigure}[]{0.32\linewidth}
        \includegraphics[scale=0.18]{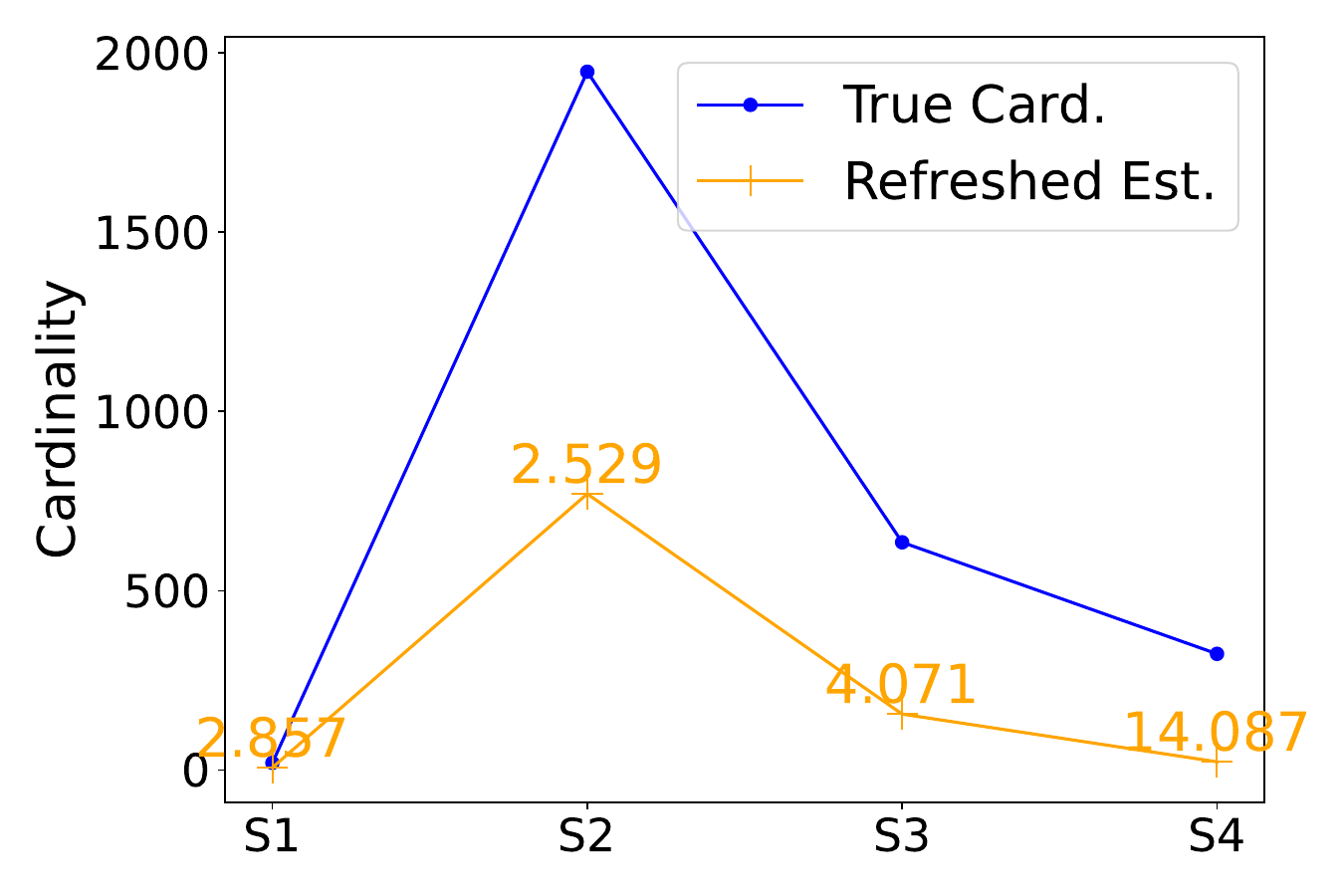}
        \caption{$\mathit{tk\_tx} \join \mathit{tx} \join a$}
    \end{subfigure}
    \\
    \begin{subfigure}[]{0.32\linewidth}
        \includegraphics[scale=0.18]{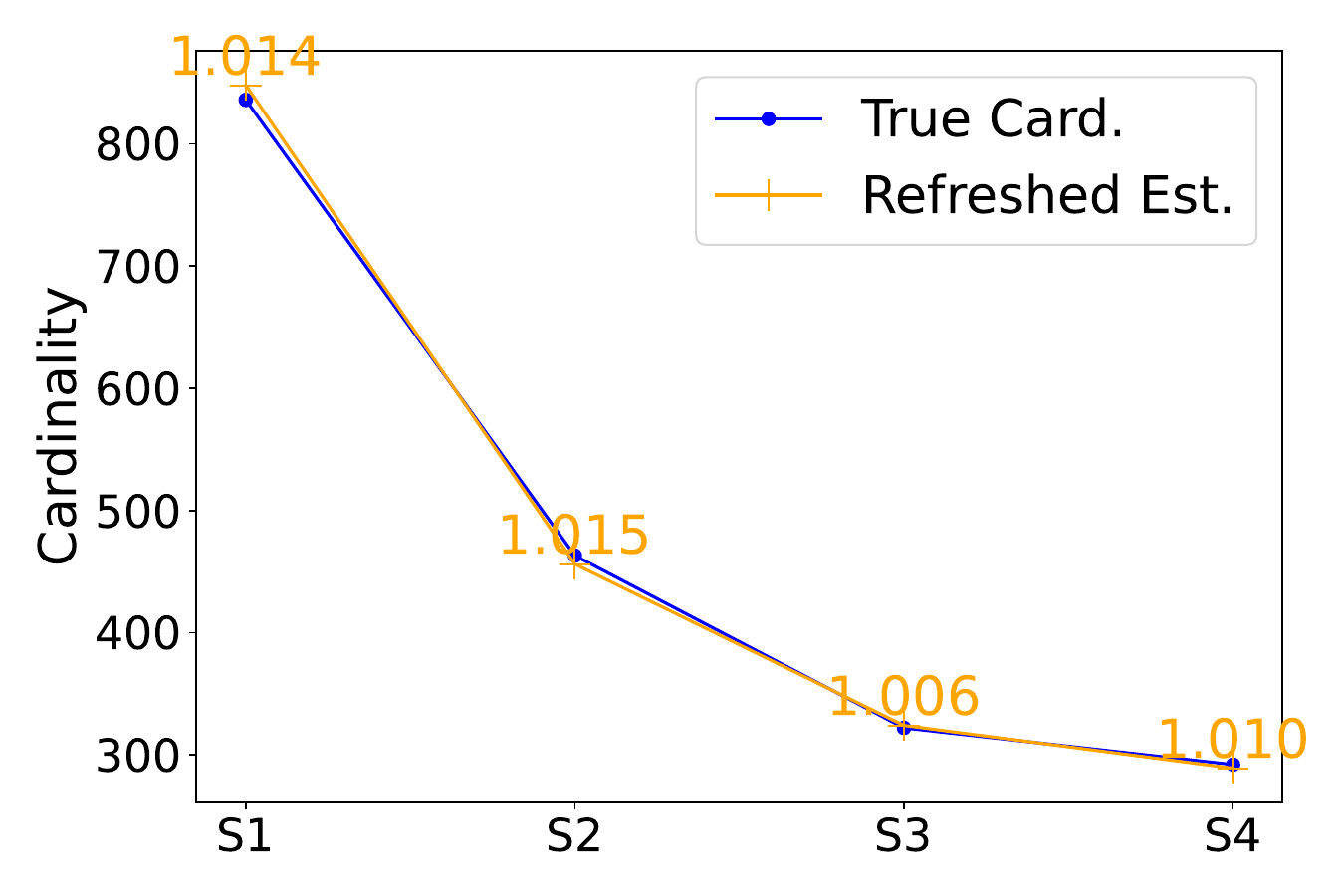}
        \caption{$\mathit{tk}$}
    \end{subfigure}
    \begin{subfigure}[]{0.32\linewidth}
        \includegraphics[scale=0.18]{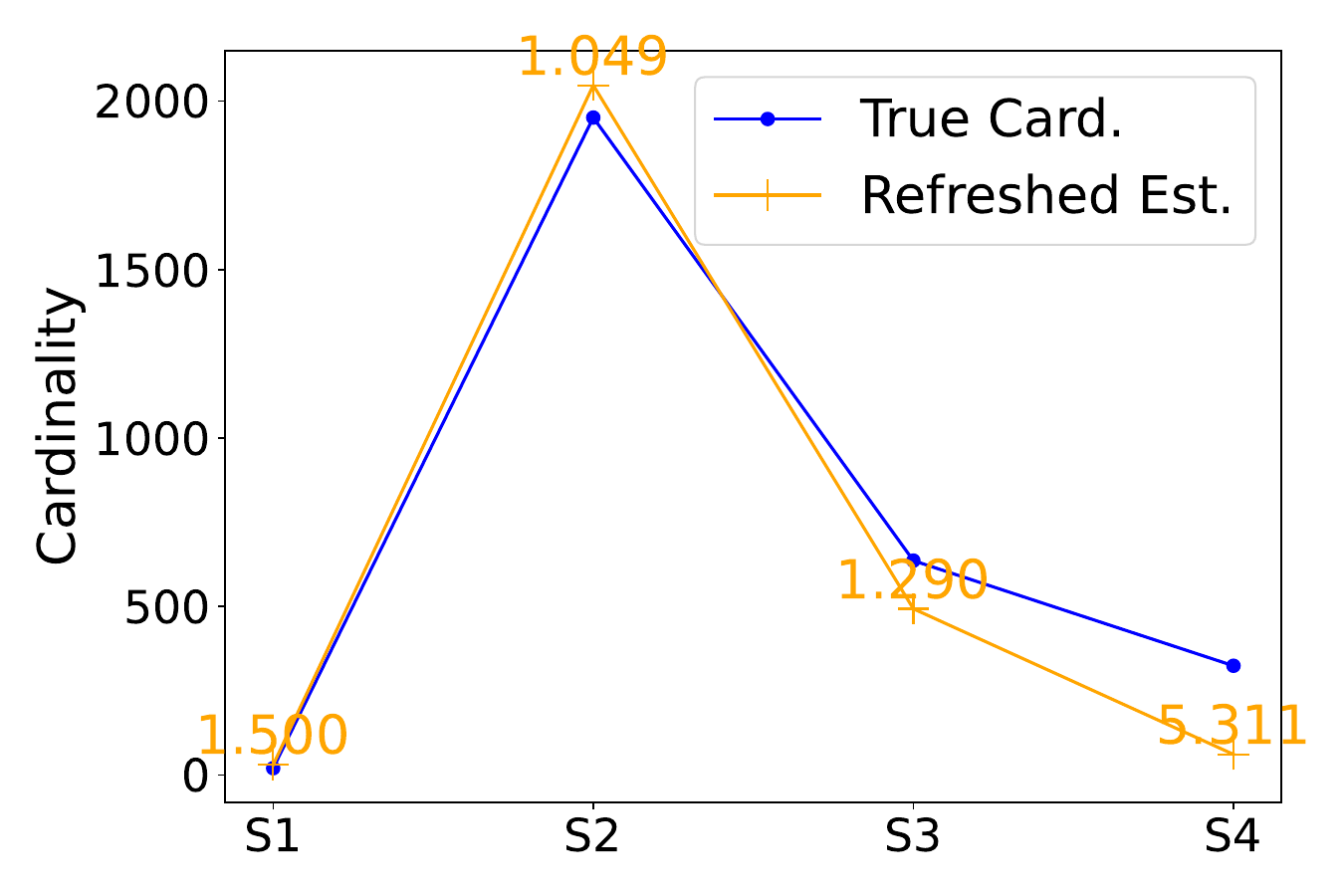}
        \caption{$\mathit{tx} \join \mathit{tk\_tx}$}
    \end{subfigure}
    \begin{subfigure}[]{0.32\linewidth}
        \includegraphics[scale=0.18]{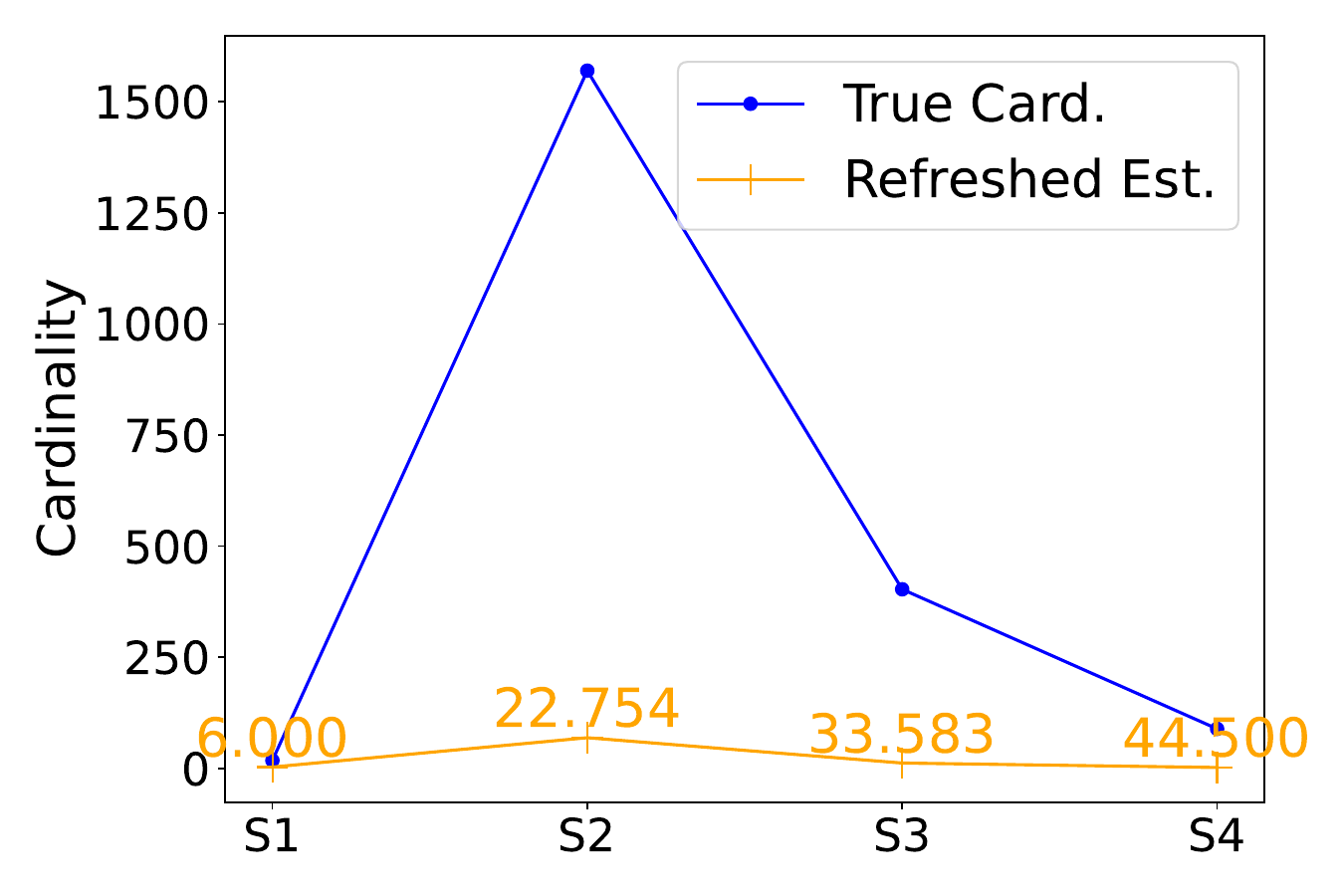}
        \caption{$\mathit{tk\_tx} \join \mathit{tx} \join c$}
    \end{subfigure}
    \\
    \begin{subfigure}[]{0.32\linewidth}
        \includegraphics[scale=0.18]{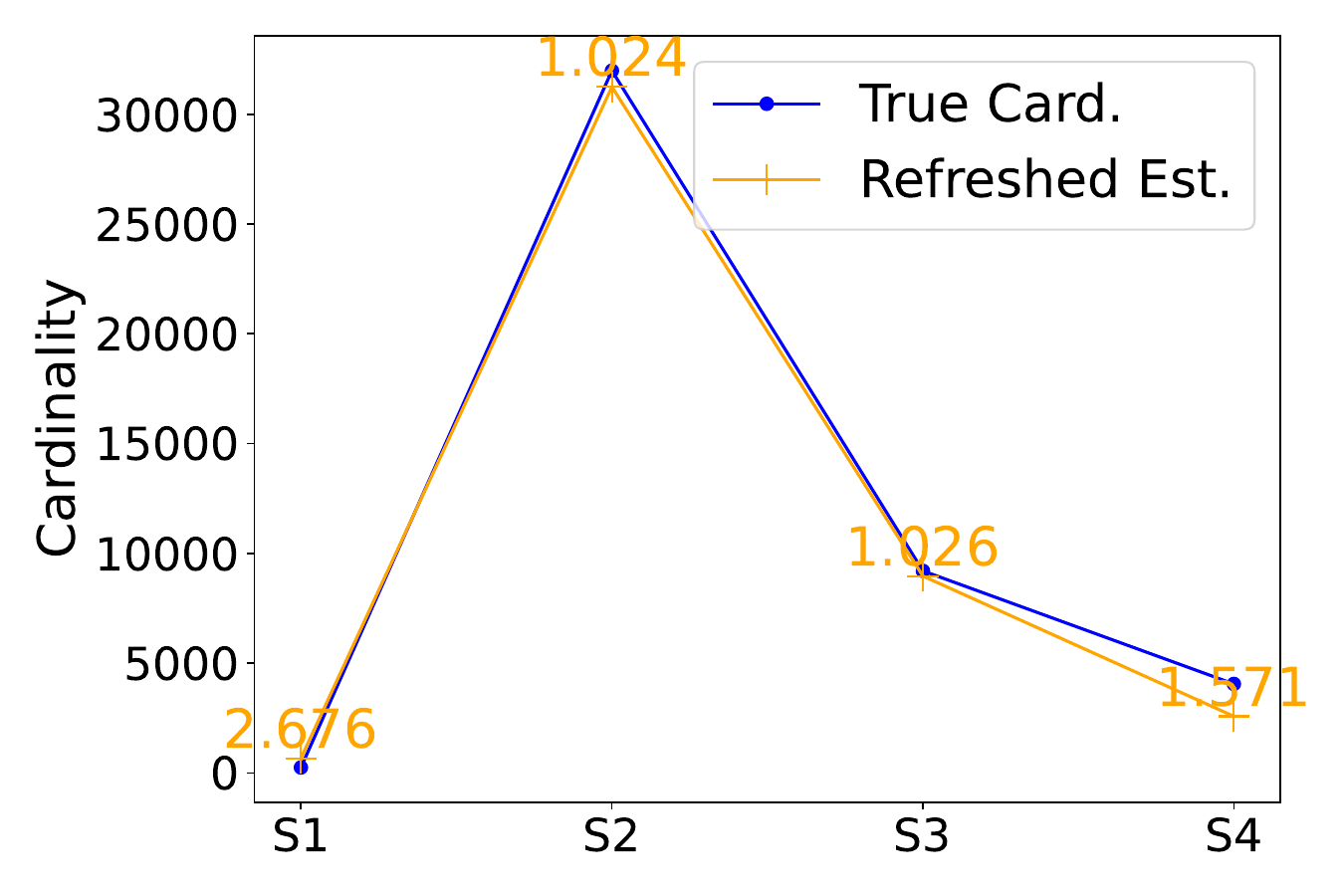}
        \caption{$\mathit{tx}$}
    \end{subfigure}
    \begin{subfigure}[]{0.32\linewidth}
        \includegraphics[scale=0.18]{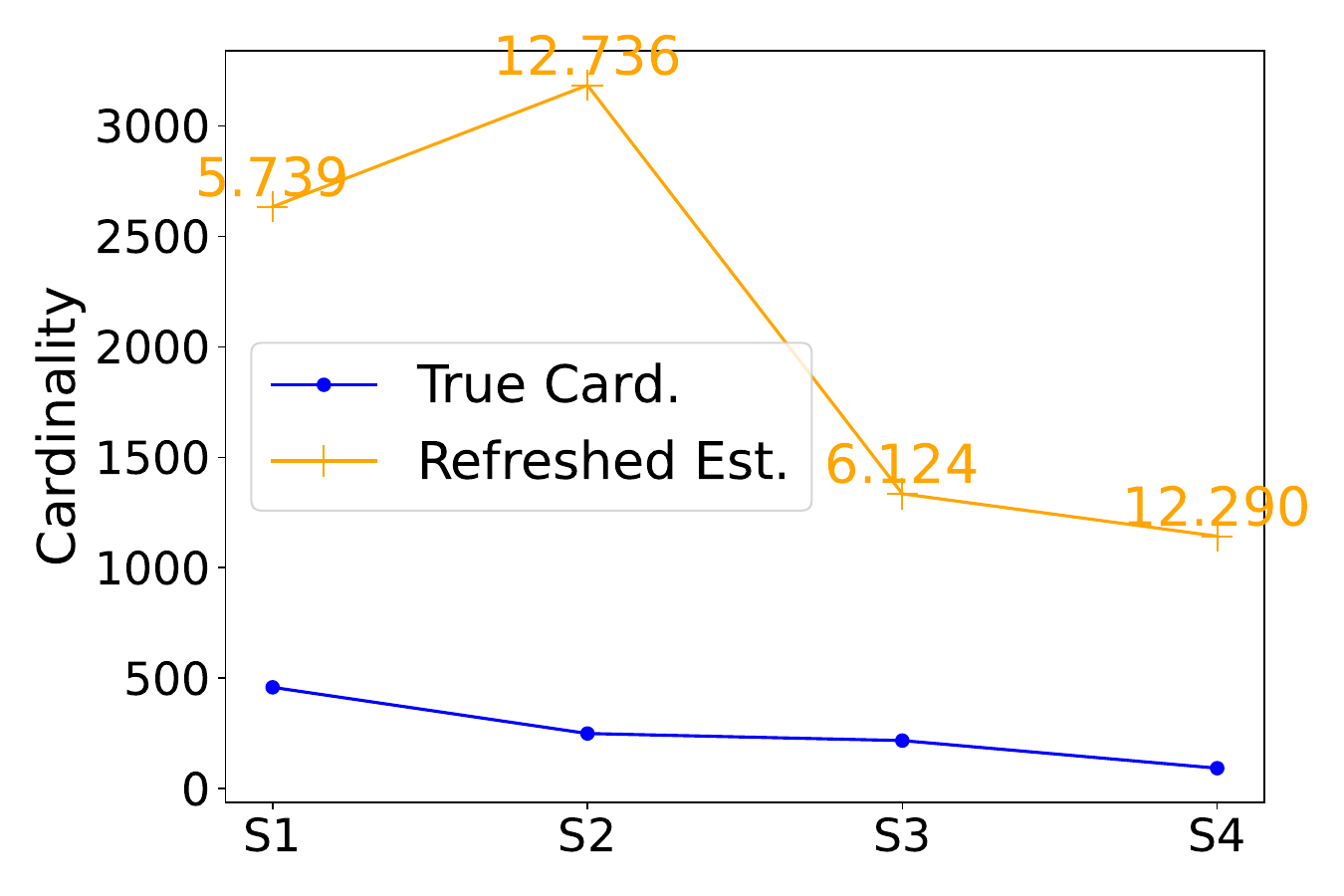}
        \caption{$\mathit{tk} \join \mathit{tk\_tx}$}
    \end{subfigure}
    \begin{subfigure}[]{0.32\linewidth}
        \includegraphics[scale=0.18]{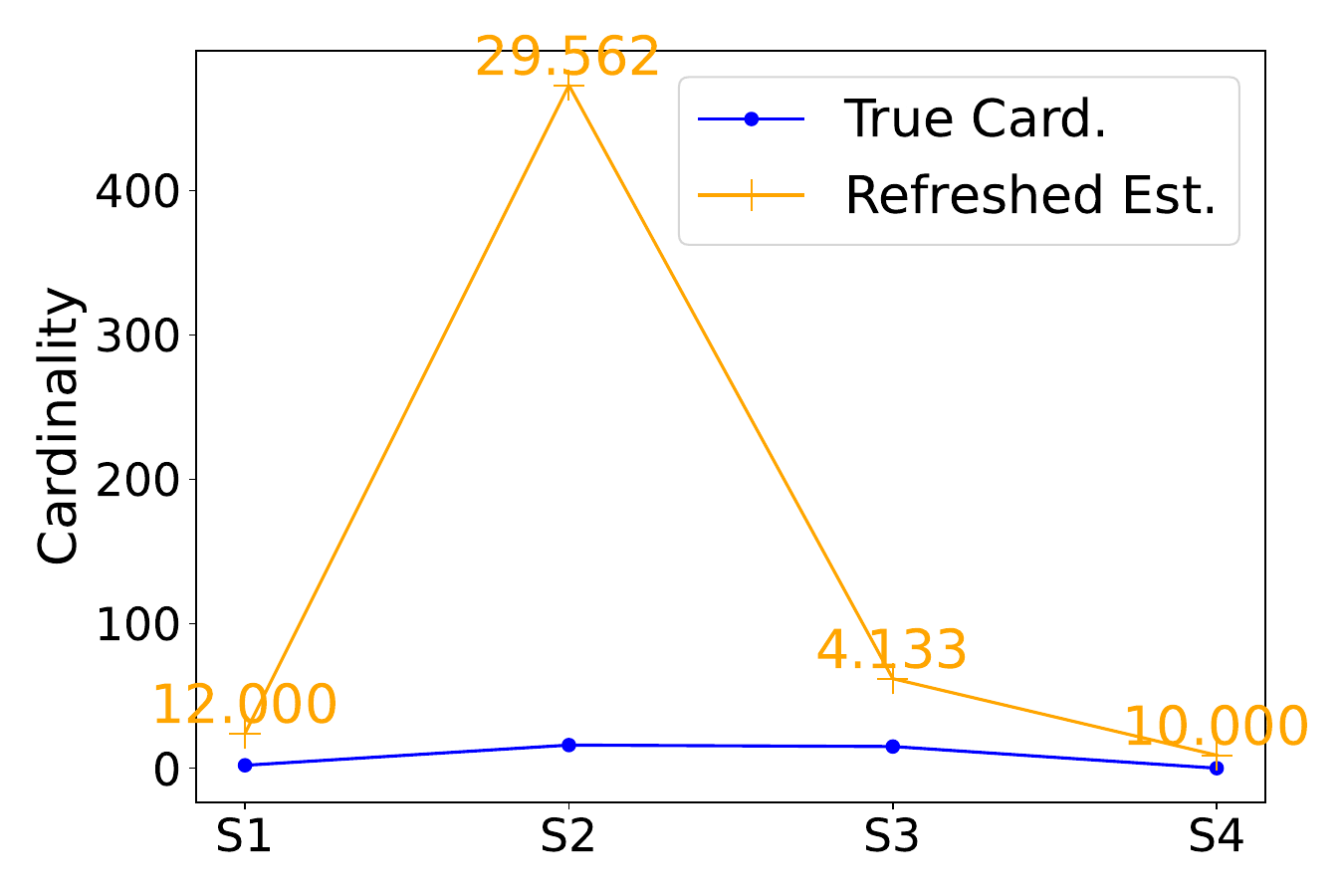}
        \caption{$\mathit{tk} \join \mathit{tk\_tx} \join \mathit{tx}$}
    \end{subfigure}
    \caption{Accuracy of cardinality estimation for Scenario~2.
    \itshape Q-errors with respect to true cardinalities are shown as labels on the lines.}
    \label{fig:qerror}
\end{figure}

For Scenario~2, the results are presented in \Cref{fig:qerror}.
For this scenario, we instruct PostgreSQL to refresh its histograms for each of the database slices.
Across the four slices $S_1$ through $S_4$, we see dramatic variations in true cardinalities for majority of the subqueries.
For example, as shown in \Cref{fig:qerror}j, the size of the \emph{Transactions} table, with the predicate $2100000 \leq \mathit{tx}.\mathit{nonce} \leq 4200000$, is $250$ for $S_1$ but $31,983$ on $S_2$.
Nonetheless, since this subquery involves only a single table, Q-errors are very small.
However, for subqueries joining more tables, Q-errors can be substantial.
Notably, in \Cref{fig:qerror}c, which is a three-way join, Q-error for 3 out of the 4 slices exceeds $10$.
Even for a two-way join in \Cref{fig:qerror}e, Q-error is $8.211$ for $S_3$.
Furthermore, although refreshing histograms in this scenario is able to pick up some trends in the variations of cardinalities across slices,
the Q-errors are much less consistent, unlike what we observed in Scenario~1, where changes between database states are smaller.
These results highlights the difficulty of consistently and accurately predicting cardinalities
in a challenging real-world dataset.

\paragraph{Example Use 2: Query Plan Selection}

After evaluating cardinaility estimation, we now examine how inaccurate estimates impact the selection of efficient query plans.
In the following, we use $P(Q, S)$ to denote the optimal plan chosen by PostgreSQL for a query $Q$ optimized on a slice $S$.
When it is clear that we are referring to a given query $Q$, we omit $Q$ from these notations. 
We measure plan performance in two ways:
\begin{itemize}
    \item Estimated plan cost: $C_e(\pi, S)$ is the cost of executing plan $\pi$ on slice $S$ as estimated by the optimizer.
    \item Actual response latency: $C_r(\pi, S)$ is the measured running time of plan $\pi$ on slice $S$, in milliseconds.
    To reduce measurement errors, we execute each plan at least 11 times and record the median.
\end{itemize}

Besides executing the plan chosen by the optimizer for a particular slice on the same slice,
we also execute it on other slices,
allowing us to evaluate how a plan performs under situations
where cardinality estimates may contain errors,
including when the underlying data has changed significantly since the last refresh of summary statistics and/or models.

We will focus on the performance of $Q_1$.
From $W_1$ to $W_{11}$ in Scenario~1, it turns out the PostgreSQL optimizer consistently picks the same execution plan.
The explanation is that although data has evolved across these database states,
the distributional changes are not significant enough to change plan selection,
even if they are significant enough to cause some large Q-errors in cardinality estimates.
This result shows that errors in cardinality estimates do not always lead to suboptimal plan selection,
and cached plans may work fine even with dynamic data updates.

However, the results for the four database slices $S_1$ to $S_4$ from Scenario~2 reveal more subtlties.
First, PostgreSQL chooses the same plan for $S_2$, $S_3$, and $S_4$, but a different plan for $S_1$.
\Cref{table:inj_cost} and \Cref{table:inj_latency} summarize the estimated plan costs ($C_e$) and actual response latencies ($C_r$), respectively.
To obtain the table entry with row heading $P(S_x)$ and column heading $S_y$,
henceforth denoted by $(P(S_x), S_y)$,
we execute on slice $S_y$ the plan optimized for slice $S_x$,
and then compute the relative speed/regression with respect to the plan optimized for $S_y$ itself,
using the ratio $C_e(P(S_x), S_y) : C_e(P(S_y), S_y)$ for \Cref{table:inj_cost} or 
$C_r(P(S_x), S_y) : C_r(P(S_y), S_y)$ for \Cref{table:inj_latency}.

\begin{table*}[!t]
    \begin{minipage}{0.49\textwidth}
    \begin{center}
        \begin{tabular}{r||c|c|c|c|}
               & $S_1$ & $S_2$ & $S_3$ & $S_4$
             \\\hline\hline
             $P(S_1)$ & - & $\downarrow1.08\times$ & $\downarrow1.75\times$ & $\downarrow2.35\times$
             \\\hline
             $P(S_2)$ & $\downarrow2.65\times$ & - & $1.00\times$ & $1.00\times$
             \\\hline
             $P(S_3)$ & $\downarrow2.65\times$ & $1.00\times$ & - & $1.00\times$
             \\\hline
             $P(S_4)$ & $\downarrow2.65\times$ & $1.00\times$ & $1.00\times$ & -
             \\\hline
        \end{tabular}
    \end{center}
    \caption{Speedup ($\uparrow$) or regression ($\downarrow$) relative to $C_e(P(S_y), S_y)$, estimated by PostgreSQL.}
    \label{table:inj_cost}
    \end{minipage}
    \hfill
    \begin{minipage}{0.49\textwidth}
    \begin{center}
        \begin{tabular}{r||c|c|c|c|}
               & $S_1$ & $S_2$ & $S_3$ & $S_4$
             \\\hline\hline
             $P(S_1)$ & - & $\downarrow2.16\times$ & $\downarrow1.42\times$ & $\downarrow1.81\times$
             \\\hline
             $P(S_2)$ & $\uparrow1.16\times$ & - & $\downarrow1.03\times$ & $1.00\times$
             \\\hline
             $P(S_3)$ & $\uparrow1.16\times$ & $\downarrow1.07\times$ & - & $1.00\times$
             \\\hline
             $P(S_4)$ & $\uparrow1.16\times$ & $\downarrow1.07\times$ & $\downarrow1.03\times$ & -
             \\\hline
        \end{tabular}
    \end{center}
    \caption{Speedup ($\uparrow$) or regression ($\downarrow$) relative to $C_r(P(S_y), S_y)$, the actual response latency.}
    \label{table:inj_latency}
    \end{minipage}
\end{table*}

For $(P(S_1), S_4)$ in \Cref{table:inj_cost}, we find a $2.35\times$ regression,
indicating that PostgreSQL considers the optimal plan chosen for $S_1$ a poor fit when running on $S_4$,
incurring a $2.35\times$ slowdown compared with the plan optimized for $S_4$ itself.
Specifically, $C_e(P(S_1), S_4) = 15701.91$ and $C_e(P(S_4), S_4) = 6689.42$.
Looking at the same cell in \Cref{table:inj_latency}, we see a $1.81\times$ slowdown,
where $C_r(P(S_1), S_4) = 32.17$ms and $C_r(P(S_4), S_4) = 17.79$ms.
We observe the same pattern in cells $(P(S_1),S_2)$ and $(P(S_1),S_3)$,
implying that $P(S_1)$ is suboptimal for the other three slices.
In these cases, PostgreSQL's cost model is effective,
since the regression in estimated costs is reflected by the regression in actual response latencies.

\begin{figure}[t]
    \begin{subfigure}[]{0.48\linewidth}
        \centering
        
        \includegraphics[scale=0.4]{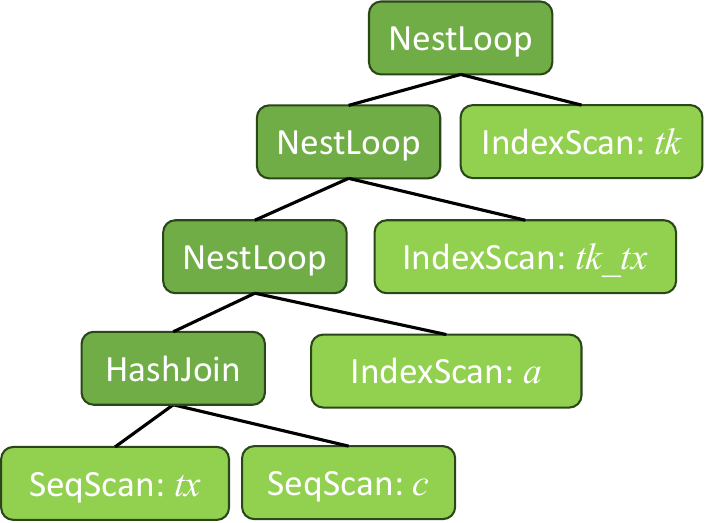}
        \caption{$P(S_1)$}
        \label{fig:plan-S1}
    \end{subfigure}
    \begin{subfigure}[]{0.48\linewidth}
        \centering
        \includegraphics[scale=0.4]{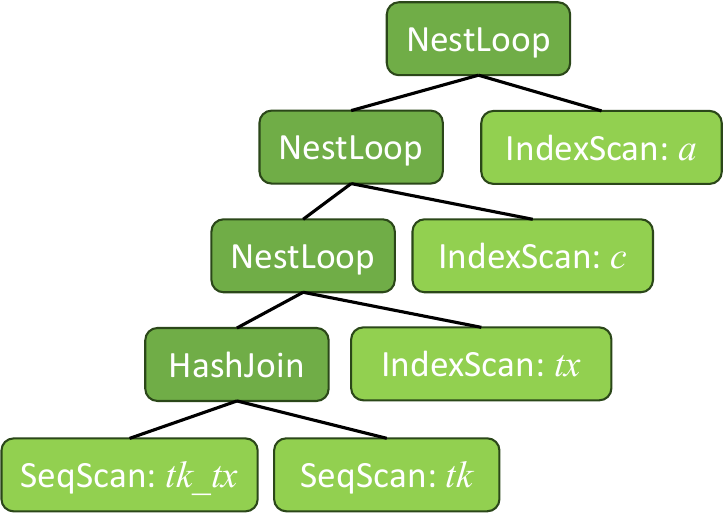}
        \caption{$P(S_2) = P(S_3) = P(S_4)$}
        \label{fig:plan-S2S3S4}
    \end{subfigure}

    \caption{Details of query plans.}
    \label{fig:plan-tree}
\end{figure}

However, we also find cases revealing inconsistency and suboptimality in the results of the PostgreSQL optimizer.
For $P(S_2) = P(S_3) = P(S_4)$ running on $S_1$, PostgreSQL expects a $2.65\times$ regression according to \Cref{table:inj_cost}.
However, actual executions achieve a $1.16\times$ speedup,
meaning that $P(S_1)$ is in fact not optimal for $S_1$.
To dig deeper, let us examine these plans in detail in \Cref{fig:plan-tree}.
%
A major reason for the poor performance of plan $P(S_1)$ on slices $S_2$, $S_3$, and $S_4$ is that the size of the table \emph{Transactions} on slice $S_2$, $S_3$ or $S_4$ is at least $10$ times larger than in slice $S_1$. Consequently, initiating a sequential scan (\texttt{SeqScan(tx)}) on \emph{Transactions} is inefficient for $S_2, S_3$ and $S_4$, particularly when there are 4 additional joins over its output.
Conversely, the advantage of $P(S_2) = P(S_3) = P(S_4)$ over $P(S_1)$ on $S_1$, which was not predicted by the optimizer,
can be attributed to PostgreSQL's overestimation of the cardinality of the subquery $(\mathit{tk\_tx} \join \mathit{tk}) \join \mathit{tx}$.
For this subquery, we observed that the cardinality estimation was over by a factor of $5.739$ on the inner join (\Cref{fig:qerror}k)
and over by $12$ when it is further joined with the third table (\Cref{fig:qerror}l).
These poor estimates cause PostgreSQL to overestimate the plan cost for $(\mathit{tk\_tx} \join \mathit{tk}) \join \mathit{tx}$ and,
consequently, the plan $P(S_2) = P(S_3) = P(S_4)$, which consists of this subquery;
PostgreSQL hence mistakenly chooses $P(S_1)$ over $P(S_2) = P(S_3) = P(S_4)$.

At least two interesting observations can be made with these results.
First, changes in data distributions have a big impact on the accuracy of cardinality estimates, which further influences plan selection.
Second, how plan selections affect query performance requires much more nuanced analysis,
as errors in cardinality and cost estimation and the heuristic nature of query optimization interact in intricate ways.
As a complex, realistic benchmark, CrypQ exposes many of such intricacies
that stress-test query optimizers and challenge traditional design decisions
in ways that benchmarks based on synthetic datasets cannot.

\section{Conclusion}
\label{sec:conclusion}
In this paper, we have introduced CrypQ, a new database benchmark that addresses limitations of current benchmarks by offering a real-world, scalable dataset along with realistic and challenging data distribution and update workload.
By leveraging the public availability, massive volume, and ever-evolving nature of data about the popular Ethereum blockchain,
CrypQ provides a myriad of opportunities for benchmarking database systems.
For example, we have demonstrated its utility in studying cardinality estimation and query plan selection in PostgreSQL.
Our query workload, currently with 10 queries of varying complexity, is very much still work in progress.
We plan to add more queries for blockchain analytics and expand it into a dynamic query workload.
We welcome community's adoption, suggestion, and contributions at \url{https://github.com/dukedb-crypq}.

\begin{credits}
\subsubsection{\ackname} This work is supported by the U.S.\ National Science Foundation awards IIS-2008107 and IIS-2402823.
\end{credits}


\bibliographystyle{splncs04}
\bibliography{bibtex}

\begin{thebibliography}{10}
\providecommand{\url}[1]{\texttt{#1}}
\providecommand{\urlprefix}{URL }
\providecommand{\doi}[1]{https://doi.org/#1}

\bibitem{our-benchmark}
{CrypQ}: A database benchmark based on dynamic, ever-evolving {Ethereum} data. \url{https://github.com/dukedb-crypq}

\bibitem{ethereum-bigquery}
{Ethereum} in {BigQuery}: a public dataset for smart contract analytics. \url{https://cloud.google.com/blog/products/data-analytics/ethereum-bigquery-public-dataset-smart-contract-analytics}

\bibitem{linear_road}
Arasu, A., Cherniack, M., Galvez, E., Maier, D., Maskey, A.S., Ryvkina, E., Stonebraker, M., Tibbetts, R.: Linear road: a stream data management benchmark. In: Proceedings of the Thirtieth International Conference on Very Large Data Bases-Volume 30. pp. 480--491 (2004)

\bibitem{ycsb}
Cooper, B.F., Silberstein, A., Tam, E., Ramakrishnan, R., Sears, R.: Benchmarking cloud serving systems with ycsb. In: Proceedings of the 1st ACM Symposium on Cloud Computing. pp. 143--154 (2010)

\bibitem{ding2021dsb}
Ding, B., Chaudhuri, S., Gehrke, J., Narasayya, V.: Dsb: A decision support benchmark for workload-driven and traditional database systems. Proceedings of the VLDB Endowment  \textbf{14}(13),  3376--3388 (2021)

\bibitem{gray1992benchmark}
Gray, J.: Benchmark handbook: for database and transaction processing systems. Morgan Kaufmann Publishers Inc. (1992)

\bibitem{smartbench}
Gupta, P., Carey, M.J., Mehrotra, S., Yus, o.: Smartbench: A benchmark for data management in smart spaces. Proceedings of the VLDB Endowment  \textbf{13}(12),  1807--1820 (2020)

\bibitem{han2021cardinality}
Han, Y., Wu, Z., Wu, P., Zhu, R., Yang, J., Tan, L.W., Zeng, K., Cong, G., Qin, Y., Pfadler, A., et~al.: Cardinality estimation in {DBMS}: A comprehensive benchmark evaluation. arXiv preprint arXiv:2109.05877  (2021)

\bibitem{hendawi2019benchmarking}
Hendawi, A., Gupta, J., Liu, J., Teredesai, A., Ramakrishnan, N., Shah, M., El-Sappagh, S., Kwak, K.S., Ali, M.: Benchmarking large-scale data management for internet of things. The Journal of Supercomputing  \textbf{75},  8207--8230 (2019)

\bibitem{espbench}
Hesse, G., Matthies, C., Perscheid, M., Uflacker, M., Plattner, H.: Espbench: The enterprise stream processing benchmark. In: Proceedings of the ACM/SPEC International Conference on Performance Engineering. pp. 201--212 (2021)

\bibitem{leis2015good}
Leis, V., Gubichev, A., Mirchev, A., Boncz, P., Kemper, A., Neumann, T.: How good are query optimizers, really? Proceedings of the {VLDB Endowment}  \textbf{9}(3),  204--215 (2015)

\bibitem{marcus2021bao}
Marcus, R., Negi, P., Mao, H., Tatbul, N., Alizadeh, M., Kraska, T.: Bao: Making learned query optimization practical. In: Proceedings of the 2021 International Conference on Management of Data. pp. 1275--1288 (2021)

\bibitem{qerror}
Moerkotte, G., Neumann, T., Steidl, G.: Preventing bad plans by bounding the impact of cardinality estimation errors. Proceedings of the VLDB Endowment  \textbf{2}(1),  982--993 (2009)

\bibitem{tpcc}
{TPC-C}: {\url{https://www.tpc.org/tpcc/}}

\bibitem{tpcds}
{TPC-DS}: {\url{https://www.tpc.org/tpcds/}}

\bibitem{xiu2024parqo}
Xiu, H., Agarwal, P.K., Yang, J.: {PARQO}: Penalty-aware robust plan selection in query optimization. Proceedings of the VLDB Endowment  \textbf{17}(13) (2024)

\end{thebibliography}

\end{document}